\RequirePackage{ifpdf}
\ifpdf 
\documentclass[pdftex]{sigma}
\else
\documentclass{sigma}
\fi

\usepackage{stmaryrd}
\usepackage[all]{xy}
\usepackage{bbm}
\usepackage{wrapfig}
\def\a{\alpha}
\def\b{\beta}

\def\d{\delta}
\def\de{\delta}

\def\g{\gamma}
\def\p{\psi}

\def\th{\theta}

\def\s{\sigma}

\def\sfrac#1#2{{\textstyle\frac{#1}{#2}}}

\def\ph{\phantom{-}}
\def\ic{{\mathrm i}}
\def\eu{{\mathrm e}}
\def\diff{{\mathrm d}}

\def\pa{\partial}
\def\>{\rangle}
\def\<{\langle}
\def\+{\dagger}
\def\ax{\a \cdot x}
\def\bx{\b \cdot x}
\def\gx{\g \cdot x}
\def\lap{\lambda_\pi}
\def\api{A_\pi^{\phantom{\top}}}
\def\apt{A_\pi^\top}

\newcommand{\R}{\mathbb R}

\newcommand{\unity}{\mathbbm{1}}
\newcommand{\mbu}{\mathbbm{u}}

\newcommand{\bfu}{\boldsymbol{u}}

\newcommand{\bfA}{\boldsymbol{A}}
\newcommand{\bff}{\boldsymbol{f}}

\newcommand{\cN}{{\cal N}}

\begin{document}

\allowdisplaybreaks

\renewcommand{\thefootnote}{$\star$}

\renewcommand{\PaperNumber}{023}

\FirstPageHeading

\ShortArticleName{$\mathcal{N}{=}4$ Multi-Particle Mechanics, WDVV Equation and Roots}

\ArticleName{$\boldsymbol{\mathcal{N}{=}4}$ Multi-Particle Mechanics,\\ WDVV Equation and Roots\footnote{This
paper is a contribution to the Proceedings of the Workshop ``Supersymmetric Quantum Mechanics and Spectral Design'' (July 18--30, 2010, Benasque, Spain). The full collection
is available at
\href{http://www.emis.de/journals/SIGMA/SUSYQM2010.html}{http://www.emis.de/journals/SIGMA/SUSYQM2010.html}}}

\Author{Olaf LECHTENFELD, Konrad SCHWERDTFEGER and Johannes TH\"URIGEN}

\AuthorNameForHeading{O.~Lechtenfeld, K.~Schwerdtfeger and J.~Th\"urigen}

\Address{Institut f\"ur Theoretische Physik, Leibniz Universit\"at Hannover,\\ Appelstrasse 2, 30167 Hannover, Germany}
\Email{\href{mailto:lechtenf@itp.uni-hannover.de}{lechtenf@itp.uni-hannover.de}, \href{mailto:k.w.s@gmx.net}{k.w.s@gmx.net},
       \href{mailto:thurigen@itp.uni-hannover.de}{thurigen@itp.uni-hannover.de}}
\URLaddress{\url{http://www.itp.uni-hannover.de/~lechtenf/}}

\ArticleDates{Received November 14, 2010, in f\/inal form February 24, 2011;  Published online March 05, 2011}

\Abstract{We review the relation of $\mathcal{N}{=}4$ superconformal multi-particle models
on the real line to the WDVV equation and an associated linear equation
for two prepoten\-tials,~$F$~and~$U$.
The superspace treatment gives another variant of the integrability problem,
which we also reformulate as a search for closed f\/lat Yang--Mills connections.
Three- and four-particle solutions are presented.
The covector ansatz turns the WDVV equation into an algebraic condition,
for which we give a formulation in terms of partial isometries.
Three ideas for classifying WDVV solutions are developed: ortho-polytopes,
hypergraphs, and matroids. Various examples and counterexamples are displayed.}


\Classification{70E55; 81Q60; 17B22; 52B40; 05C65}

\renewcommand{\thefootnote}{\arabic{footnote}}
\setcounter{footnote}{0}

\section{Introduction}

Over the past decade, there has been substantial progress in the construction
of $\cN{=} 4$ superconformal multi-particle mechanics (in one space dimension)
\cite{tolik,wyl,bgk,bgl,glp1,glp2,bks,fil,klp,lp}.
In 2004 a~deep connection between these physical systems and the so-called
WDVV equation~\cite{w,dvv} was discovered~\cite{bgl}, relating Calogero-type
models with $D(2,1;\alpha)$ superconformal symmetry to a~branch of mathematics
concerned with solving this equation~\cite{dubrovin,margra,ves,strachan,chaves,feives1,feives2}.
Here, we describe physicists' attempts to take advantage of the
mathematical literature on this subject and to develop it further towards
constructing and classifying such multi-particle models.

There exist dif\/ferent versions of the WDVV equation in the literature,
so let us be more specif\/ic. Originally, the WDVV equation appeared as a~consistency relation in topological f\/ield theory, where the puncture operator
singles out one of the coordinates, so that the as\-so\-cia\-ted Frobenius algebra
is unital and carries a constant metric~\cite{w,dvv,dubrovin}.
A few years later, a~more general form of the WDVV equation appeared as a~condition on the prepotential~$\widetilde{F}$ of Seiberg--Witten theory
(four-dimensional $\cN{=} 2$ super Yang--Mills theory)~\cite{bonmat,mmm,mironov}.
Here, the distinguished coordinate is absent, and so the Frobenius structure
constants do not lead to a~natural metric. This so-called generalized WDVV
equation takes the form
\begin{gather} \label{wdvvgeneral}
\widetilde\bfA_i \widetilde\bfA_k^{-1}\widetilde\bfA_j  =
\widetilde\bfA_j \widetilde\bfA_k^{-1}\widetilde\bfA_i
\qquad \forall\, i,j,k=1,\ldots,n
\end{gather}
for a collection of $n \times n$ matrix functions
$(\widetilde\bfA_i)_{\ell m}=-\pa_i\pa_\ell\pa_m \widetilde{F}$.
However, any invertible linear combination of these matrices yields an
admissible metric
\begin{gather} \label{etametric}
\boldsymbol{\eta}= \sum_i \eta^i\widetilde\bfA_i
\qquad\textrm{so that}\qquad
\widetilde\bfA_i \boldsymbol{\eta}^{-1}\widetilde\bfA_j  =
\widetilde\bfA_j \boldsymbol{\eta}^{-1}\widetilde\bfA_i  .
\end{gather}
It is easy to see that this metric can be absorbed
in a redef\/inition~\cite{margra,ves},
\[ 
\bfA_i  :=  \boldsymbol{\eta}^{-1} \widetilde\bfA_i
\quad\longrightarrow\quad
\bigl[\bfA_i , \bfA_j\bigr] =0
\qquad\textrm{with}\quad
{\textstyle\sum_i}\eta^i\bfA_i=\unity  ,
\]
giving a formulation equivalent to~(\ref{wdvvgeneral}).
For a constant metric,
e.g.\ $\boldsymbol{\eta}=\widetilde\bfA_1=\textrm{const}$,
we fall back to the more special case which arose in topological f\/ield theory.
Since 1999, Veselov and collaborators have been constructing particular
solutions to~(\ref{etametric}), introducing so-called $\vee$-systems and
featuring a {\it constant\/} metric
$\boldsymbol{\eta}=\sum_ix^i\widetilde\bfA_i$
\cite{ves,chaves,feives1,feives2}.

In comparison, $\cN{=} 4$ supersymmetric multi-particle models are determined
by two prepotentials, $F$ and~$U$, the f\/irst of which is subject to the
generalized WDVV equation~\cite{bgl}. Here, the conformal invariance
imposes a supplementary condition on our matrices, which
amounts to choosing the {\it Euclidean\/} metric
\[
\sum_i x^i \widetilde\bfA_i  = \unity
\qquad\longrightarrow\qquad \widetilde\bfA_i=\bfA_i  ,
\]
and so one may drop the label `generalized'.
The map to Veselov's formulation is achieved by a~linear coordinate change,
$x^i\to x^j M_j^{\ i}$   with
$\boldsymbol{\eta}=\boldsymbol{M}\boldsymbol{M}^\perp$~\cite{letalk}.

The goal of the paper is fourfold.
First, we would like to review the appearance of the WDVV equation in the
construction of one-dimensional multi-particle models with $su(1,1|2)$
symmetry~\cite{bgl,glp1,glp2,bks,fil,klp,lp}.
In particular, we draw the attention of the mathematical
readers to the {\it second\/} prepotential~$U$, which enlarges the WDVV
structure in a canonical fashion, and to the superspace formulation, which
yields an alternative formulation of the integrability condition.
Second, we plan to provide some explicit three- and four-particle
examples for the physical model builders.
Third, we intend to rewrite Veselov's notion of $\vee$-systems in a manner
we hope is more accessible to physicists, using the notion of partial isometry
and providing further examples.
Fourth, we want to advertize some novel attempts to attack the classif\/ication
problem for the WDVV equation. The standard ansatz for the propotential employs
a collection of covectors, which are subject to intricate algebraic conditions.
These relations may be visualized in terms of certain polytopes, or in terms
of hypergraphs, or by a particular kind of matroid. Neither of these concepts
is fully satisfactory; the classif\/ication problem remains open. However, for
low dimension and a small number of covectors they can solve the problem.

\section{Conformal quantum mechanics: Calogero system}

As a warm-up, we introduce
$n + 1$ identical particles with unit mass, moving on the real line,
with coordinates $x^i$ and momenta $p_i$, where $i=1,2,\ldots,n + 1$,
and def\/ine their dynamics by the Hamiltonian\footnote{Equivalently, it describes a single particle moving in $\R^{n+1}$
under the inf\/luence of the external potential $V_B$.}
\begin{gather}\label{h}
H =\sfrac{1}{2} p_i p_i  +  V_B \big(x^1, \dots, x^{n+1}\big)   .
\end{gather}
For the quantum theory, we impose the canonical commutation relations
($\hbar = 1$)
\[
[x^i, p_j] =\ic {\de_j}^i   .
\]
Together with the dilatation and conformal boost generators
\[
D =-\sfrac{1}{4} \big(x^i p_i +p_i x^i\big) \qquad{\rm and}\qquad
K = \sfrac{1}{2} x^i x^i   ,
\]
the Hamiltonian~(\ref{h}) spans the conformal algebra $so(2,1)$
in $1 + 0$ dimensions,
\begin{gather*}
[D,H] =-\ic H  ,\qquad
[H,K] =2\ic D  ,\qquad
[D,K] =\ic K   ,
\end{gather*}
if and only if $(x^i\pa_i+2) V_B=0$, i.e.\ the potential is homogeneous
of degree~$-2$.
If one further demands permutation and translation invariance
and allows only two-body forces, one ends up with the Calogero model,
\[
V_B  = \sum_{i<j} \frac{g^2}{(x^i-x^j)^2}  .
\]

\section[$\cN = 4$ superconformal extension: $su(1,1|2)$ algebra]{$\boldsymbol{\cN{=} 4}$ superconformal extension: $\boldsymbol{su(1,1|2)}$ algebra}

Our goal is to $\cN{=} 4$ supersymmetrize conformal multi-particle mechanics.
The most general $\cN{=} 4$ extension of $so(2,1)$ is the superalgebra
$D(2,1;\a)$, but here we specialize to
$D(2,1;0)\simeq su(1,1|2)\niplus su(2)$.
Further, we break the outer $su(2)$ to $u(1)$ by allowing for
a central charge~$C$. The set of generators then gets extended
\cite{ioko}
\[
(H,D,K)\ \to\ (H,D,K,Q_\a,S_\a,J_a,C) \qquad\textrm{with}\qquad
\a=1,2 \quad\textrm{and}\quad a=1,2,3
\]
and hermiticity properties
${(Q_\a)}^\+={\bar Q}^\a$ and ${(S_\a)}^\+={\bar S}^\a$.

The nonvanishing (anti)commutators of $su(1,1|2)$ read
\begin{alignat*}{3}
& [D,H]=-\ic H,\qquad  && [H,K]=2\ic D, & \nonumber\\
& [D,K]=+\ic K, \qquad  && [J_a,J_b]=\ic \epsilon_{abc} J_c, & \nonumber\\
& \{ Q_\a, \bar Q^\b \}=2 H {\d_\a}^\b,\qquad  && \{ Q_\a, \bar S^\b \}= +2\ic {{(\s_a)}_\a}^\b J_a-2 D {\d_\a}^\b-\,\ic C {\d_\a}^\b,& \nonumber\\
& \{ S_\a, \bar S^\b \}\ =2 K {\d_\a}^\b,\qquad  && \{ \bar Q^\a, S_\b \}= -\,2\ic {{(\s_a)}_\b}^\a J_a-2 D {\d_\b}^\a+\ic C {\d_\b}^\a,& \nonumber\\
& [D,Q_\a] = -\sfrac{\ic}{2} Q_\a,\qquad  && [D,S_\a] =+\sfrac{\ic}{2} S_\a,& \nonumber\\
& [K,Q_\a] =+\ic S_\a,\qquad  && [H,S_\a] =-\ic Q_\a, &\nonumber\\
& [J_a,Q_\a] =-\sfrac{1}{2} {{(\s_a)}_\a}^\b Q_\b, \qquad && [J_a,S_\a] =-\sfrac{1}{2} {{(\s_a)}_\a}^\b S_\b,& \nonumber\\
& [D,\bar Q^\a] =-\sfrac{\ic}{2} \bar Q^\a,\qquad  && [D,\bar S^\a] =+\sfrac{\ic}{2} \bar S^\a, & \nonumber\\
& [K,\bar Q^\a] =+\ic \bar S^\a, \qquad   && [H,\bar S^\a] =-\ic \bar Q^\a,& \nonumber\\
& [J_a,\bar Q^\a] =\sfrac{1}{2} \bar Q^\b {{(\s_a)}_\b}^\a, \qquad  && [J_a,\bar S^\a] =\sfrac{1}{2} \bar S^\b {{(\s_a)}_\b}^\a . &
\end{alignat*}

To realize this algebra on the $(n+1)$-particle state space,
we must enlarge the latter by adding Grassmann-odd degrees of freedom,
$\psi^i_\a$ and $\bar\psi^{i\a}={\psi^i_\a}^\+$,
with $i=1,\dots,n+1$ and $\a=1,2$, and subject them to canonical
anticommutation relations,
\[
\{\p^i_\a, \p^j_\b \}=0 , \qquad
\{ {\bar\p}^{i\a}, {\bar\p}^{j\b} \}=0 \qquad \mbox{and}\qquad
\{\p^i_\a, {\bar\p}^{j\b} \}= {\d_\a}^\b \d^{ij}  .
\]
In the absence of a potential (subscript `0'), the generators are given
by the bilinears
\begin{gather*} 
{Q_0}_\a=p_i \p^i_\a , \qquad
\bar Q_0^\a=p_i \bar\p^{i\a} \qquad \mbox{and}\qquad
{S_0}_\a=x^i \p^i_\a , \qquad
\bar S_0^\a=x^i \bar\p^{i\a}  ,
 \\
H_0= \sfrac12 p_i p_i ,\qquad
D_0= -\sfrac{1}{4}(x^i p_i +p_i x^i) ,\qquad
K_0= \sfrac12 x^i x^i ,\qquad
{J_0}_a = \sfrac{1}{2} \bar\p^{i\a} {{(\s_a)}_\a}^\b \p^i_\b  ,
\end{gather*}
where $\s_a$ denote the Pauli matrices. Surprisingly however,
the free generators fail to obey the $su(1,1|2)$ algebra, and
interactions are mandatory! The minimal deformation touches only
the supercharge and the Hamiltonian,{\samepage
\begin{gather*} 
Q_\a  = Q_{0\a} -\ic\,[S_{0\a},V]   ,\qquad
\bar Q^\a  = \bar Q^\a_0 - \ic [\bar S^\a_0,V]
\qquad \mbox{and}\qquad H  = H_0 + V  ,
\end{gather*}
keeping $S=S_0$, $\bar{S}=\bar{S}_0$, $D=D_0$, $K=K_0$ and $J=J_0$.}

Being a Grassmann-even function of $\p$, $\bar\p$ and $x$,
the potential~$V$ may be expanded in even powers of the fermionic variables.
It turns out that we must go to fourth order for closing the algebra,
i.e.~\cite{wyl,bgl,glp2}
\begin{gather}
\label{ans}
V  = V_{\textrm{B}}(x)  -
U_{ij}(x) \langle \p^i_\a {\bar\p}^{j\a} \rangle  +
\sfrac14 F_{ijkl}(x) \langle\p^i_\a\p^{j\a}\bar\p^{k\b}\bar\p^l_\b\rangle  ,
\end{gather}
where the angle brackets $\langle\cdots\rangle$ denote symmetric (or Weyl)
ordering. The functions $U_{ij}$ and~$F_{ijkl}$ are totally symmetric
in their indices and homogeneous of degree~$-2$ in $\{x^1,\ldots,x^{n+1}\}$.
For completeness, we also give the interacting supercharge,
\begin{gather}\label{Qform}
Q_\a  = \bigl(p_j-\ic x^i U_{ij}(x)\bigr) \p_\a^j  -
\sfrac{\ic}{2} x^i F_{ijkl}(x) \<\p^j_\b \p^{k\b}\bar\p^l_\a\>  .
\end{gather}

\section[The structure equations for $(F,U)$: WDVV, Killing, inhomogeneity]{The structure equations for $\boldsymbol{(F,U)}$:\\ WDVV, Killing, inhomogeneity}

Inserting the minimal ansatz~(\ref{ans}) for $V$ into the $su(1,1|2)$ algebra
and demanding closure, one f\/inds that
\begin{gather*}
U_{ij}  = \pa_i\pa_j U \qquad \mbox{and} \qquad  F_{ijkl}  = \pa_i\pa_j\pa_k\pa_l F
\end{gather*}
are determined by two scalar prepotentials $U$ and $F$, which are
subject to so-called {\it structure equations\/}~\cite{wyl,bgl,glp2},
\begin{alignat}{3}
& (\pa_i\pa_k\pa_p F)(\pa_p\pa_l\pa_j F)-
(\pa_i\pa_l\pa_p F)(\pa_p\pa_k\pa_j F) =0,
\qquad  &&
x^i \partial_i \partial_j \partial_k F=-\d_{jk}  ,&
\label{w1} \\
&
\pa_i\pa_j U -(\pa_i\pa_j\pa_k F) \pa_k U =0,
\qquad && x^i \pa_i U =-C  .
\label{w2}
\end{alignat}
The left equations (\ref{w1}a) and (\ref{w2}a) are homogeneous quadratic in~$F$
(known as the WDVV equation)~\cite{w,dvv} and homogeneous linear in~$U$
(a~type of Killing equation).
The right equations~(\ref{w1}b) and~(\ref{w2}b) introduce well-def\/ined
inhomogeneities, so that the prepotential must be of the form
\begin{gather}\label{com}
F  = -\sfrac12 x^2 \ln x + F_{\textrm{hom}} \qquad \mbox{and}\qquad U  = -C \ln x + U_{\textrm{hom}}
\end{gather}
with $F_{\textrm{hom}}$ of degree $-2$ and $U_{\textrm{hom}}$ of degree~$0$
in~$x$. This also shows the redundancies
\[
U  \simeq  U+\textrm{const} \qquad \mbox{and}\qquad
F  \simeq  F+\textrm{quadratic polynomial}  ,
\]
which for $F$ is also apparent in the twice-integrated form of (\ref{w1}b),
\[
(x^i\pa_i - 2) F  = -\sfrac12 x^ix^i  .
\]

It is convenient to separate the center-of-mass dynamics from the
relative particle motion, since the two decouple in all equations.
The center-of-mass motion is already nonlinear but explicitly solved by
(\ref{com}) without homogeneous terms (the central charge is additive).
In new relative-motion coordinates, which again we name~$x^i$ but with
$i=1,2,\ldots,n$, the conf\/iguration space is reduced to $\R^n$.
The Killing-type equation~(\ref{w2}a) implies, as its compatibility condition,
the WDVV equation~(\ref{w1}a) contracted with~$\pa_j U$. Furthermore, the
contraction of (\ref{w1}a) with $x^i$ is trivially valid,
thanks to~(\ref{w1}b). This ef\/fectively projects the WDVV equation to~$n - 1$
dimensions. Since its symmetry is that of the Riemann tensor, it comprises
as many independent equations, namely $\sfrac1{12}n(n - 1)^2(n - 2)$ in number.
In particular, (\ref{w1}a) is empty for up to three particles and a single
condition for four particles.

The leading part of the potential is also determined by $U$ and~$F$,\footnote{Here and later, we sometimes reinstate $\hbar$ to ease the interpretation.}
\begin{gather*} 
V_{\textrm{B}}  = \sfrac12 (\pa_iU)(\pa_iU)   +
\sfrac{\hbar^2}{8} (\pa_i\pa_j\pa_kF)(\pa_i\pa_j\pa_k F)   >   0 ,
\end{gather*}
and the expressions in (\ref{Qform}) simplify to
\[
x^i F_{ijkl}  = -\pa_j\pa_k\pa_lF \qquad \mbox{and}\qquad  x^i U_{ij}  = -\pa_j U  .
\]
Therefore, f\/inding a pair $(F,U)$ amounts to def\/ining an $su(1,1|2)$ invariant
$(n + 1)$-particle model. For more than three particles, however, this is
a dif\/f\/icult task, and very little is known about the space of solutions.

\section[Superspace approach: inertial coordinates in $R^{n+1}$]{Superspace approach: inertial coordinates in $\boldsymbol{\R^{n+1}}$}

When analyzing supersymmetric systems, it is often a good idea to employ
superspace methods. This is also possible for the case at hand, where the
construction of a classical Lagrangian seems straightforward in $\cN{=} 4$
superspace~\cite{leva,ikl1,deiv,ivle,bekr}.

For each particle, we introduce a standard untwisted $\cN{=} 4$ superf\/ield
\begin{gather*}
\mbu^A(t,\th^a,\bar\th_a)  = u^A(t)+O(\th,\bar\th)
\qquad\textrm{with}\quad A=1,\ldots,n + 1  ,
\end{gather*}
obeying the constraints\footnote{The constants $g^A$ can be SU(2)-rotated into the constraints,
so that   $D^2\mbu^A=\ic g^a\!=\!-\overline{D}^2\mbu^A$
but   $[D^a\!,\overline{D}_a] \mbu^A=0$.}
\[
D^2\mbu^A=0=\overline{D}^2\mbu^A
\qquad\longrightarrow\qquad
\pa_t [D^a,\overline{D}_a] \mbu^A= 0
\qquad\longrightarrow\qquad
[D^a,\overline{D}_a] \mbu^A=2 g^A
\]
with constants $g^A$, which will turn out to be the coupling parameters.
The general $\cN{=} 4$ superconformal action for these f\/ields takes the form\footnote{Subscripts on $G$ denote derivatives with respect to $u$,
i.e.~$G_A=\pa G/\pa u^A$ etc.}
\[
S =-\int \diff t \diff^2\th \diff^2\bar\th   G(\mbu)
 = \sfrac12\int \diff t
\left[ G_{AB}(u) \dot u^A \dot u^B - G_{AB}(u) g^A g^B +
\textrm{fermions}\right]
\]
already written in~\cite{tolik},
with a superpotential $G(u)$ subject to the conformal invariance condition
\[
G-G_A u^A =\sfrac12 c_A u^A
\]
for arbitrary constants $c_A$, so that it is of the form
$G=-\sfrac12 c u\ln u+\textrm{terms of degree one}$.

Generically, such sigma-model-type actions do not admit a multi-particle
interpretation, however, unless the target space is f\/lat. This requirement
imposes a nontrivial condition on the target-space metric~$G_{AB}(u)$~\cite{klp},
\[
\textrm{Riemann}(G_{AB}) =0 \qquad\longleftrightarrow\qquad
G_{A[BX} G^{XY} G_{YC]D} =0   .
\]
Equivalently, there must exist so-called {\it inertial coordinates\/}~$x^i$,
with $i=1,2,\ldots,n + 1$, such that
\[
S  = \int \diff t \left[ \sfrac12\delta_{ij} \dot x^i \dot x^j
- V_{\textrm{B}}^{\textrm{cl}}(x) + \textrm{fermions} \right]  .
\]
The goal is, therefore, to f\/ind admissible functions   $u^A=u^A(x)$
and compute the corresponding $G$ and $V_{\textrm{B}}^{\textrm{cl}}$.
The above f\/latness requirement leads to a specif\/ic integrability condition
for   $u^A_{\ i}:=\pa_iu^A$, namely
\begin{gather}\label{int1}
\frac{\pa x^i}{\pa u^A}\bigl(u(x)\bigr)  \equiv
\left( (u^\bullet_{\ \bullet})^{-1}\right)^i_{\ A}  =:
w_{A,i}  \ {\buildrel !\over =}  \
\pa_i w_A  \equiv  \frac{\pa w_A}{\pa x^i}(x)   ,
\end{gather}
which says that the transpose of the inverse Jacobian for $u\to x$ is
again a Jacobian for a map $w\to x$. This def\/ines a set of functions $w_A(x)$
dual to $u^a(x)$, in the sense that their Jacobians are inverses~\cite{klp},
\[
w_{A,i} u^B_{\ i} =\de_A^{\ B} \qquad\longleftrightarrow\qquad
w_{A,i} u^A_{\ j} =\de_{ij}  .
\]
Equivalent versions of the integrability condition~(\ref{int1}) are~\cite{klp}
\begin{gather}
u^{[A}_{\ \ i} \pa_j u^{B]}_{\ i} =0 \qquad\longleftrightarrow\qquad
w_{[A,i} \pa_j w_{B],i} =0   ,\label{int2}\\
  f_{ijk}  :=  -w_{A,i} \pa_k u^A_{\ j}
\qquad\textrm{is totally symmetric}   ,\nonumber \\ 
  f_{ijk} =\pa_i\pa_j\pa_k F \qquad \mbox{and}\qquad
f_{im[k} f_{l]mj} = 0  ,\nonumber 
\end{gather}
which includes the WDVV equation for~$F$.
In contrast, there is no formulation purely in terms of~$U$.

Conformal invariance restricts $u^A$ to be homogeneous quadratic in~$x$,
hence $w_A$ to be homogeneous of degree zero (including logarithms!),
thus $f_{ijk}$ is of degree~$-1$.
The second prepotential~$U$ is also determined by $u^A(x)$ via
\[
U(x)  = -g^A w_A(x) \qquad\textrm{so that}\qquad
C =-x^i\pa_i U =c_Ag^A ,
\]
and automatically fulf\/ills the Killing-type equation~(\ref{w2}a).
The classical bosonic potential then reads
\[
V_{\textrm{B}}^{\textrm{cl}} = \sfrac12 (\pa_iU)(\pa_iU)
 = \sfrac12 g^A g^B w_{A,i} w_{B,i}  .
\]

Finally, for the superpotential $G(u)$ the integrability condition becomes
\[
u^A_{\ i}u^B_{\ j} G_{AB} = -\de_{ij} \qquad\longleftrightarrow\qquad
G_{AB} = -w_{A,i} w_{B,i} = -\pa_A w_B = -\pa_B w_A  ,
\]
so that, up to an irrelevant $u$-linear shift of~$G$,
\[
w_A = -G_A \qquad\longleftrightarrow\qquad G  = -u^A w_A  ,
\]
and we have
\[
U = g^A G_A \qquad \mbox{and} \qquad
f_{ijk}= -\sfrac12 u^A_{\ i}u^B_{\ j}u^C_{\ k} G_{ABC}
\qquad\longrightarrow\qquad
V_{\textrm{B}}^{\textrm{cl}}= -\sfrac12 g^A g^B G_{AB}  .
\]
However, knowing the superpotential does not suf\/f\/ice: the relation between
$x^i$ and $u^A$ is needed to determine $U(x)$ and $F(x)$.
On the other hand, if a solution~$F$ to the WDVV equation can be found,
this problem reduces to a linear one~\cite{klp}:
\begin{gather}\label{ulin}
u^A_{\ ij} + f_{ijk} u^A_{\ k} = 0 \qquad\textrm{or}\qquad
w_{A,ij} - f_{ijk} w_{A,k} = 0  ,\qquad\textrm{with}\quad
f_{ijk}=\pa_i\pa_j\pa_k F  .
\end{gather}
Finally, we remark again that the center-of-mass degree of freedom
can be decoupled, so that all indices may run form~1 to~$n$ only.

\section[Structural similarity to closed flat Yang-Mills connections]{Structural similarity to closed f\/lat Yang--Mills connections}

It is instructive to rewrite our integrability problem in terms of
$n \times n$-matrix-valued dif\/ferential forms, in a compact formulation
closer to Yang--Mills theory. To this end, we def\/ine
\begin{gather}\label{matrixdefs}
\bigl(u^A_{\ i}\bigr)  :=  \bfu  ,\qquad
\bigl(-\pa_i\pa_jF\bigr)     :=  \bff \qquad \mbox{and} \qquad
\bigl(-\pa_i\pa_j\pa_kF \diff x^k\bigr)  :=  \bfA\ =\bfA_k\diff x^k  .
\end{gather}
Since   $\bfA_k=\pa_k\bff$   and   $\pa_i\pa_j\pa_kF=w_{A,i} \pa_k u^A_{\ j}$,
we have
\[
\bfA=\diff\bff \quad\rightarrow\quad \diff\bfA=0 \qquad \mbox{and} \qquad
\bfA=\bfu^{-1}\diff\bfu \quad\rightarrow\quad
\diff\bfA+\bfA{\wedge}\bfA= 0  ,
\]
from which we learn that
\begin{gather} \label{int5}
0 = \bfA\wedge\bfA =
\sfrac12\diff [\bff , \diff\bff]=
-\diff\bfu^{-1}\wedge\diff\bfu  ,
\end{gather}
which is nothing but the WDVV equation again.
Hence, we are looking for connections~$\bfA$ which are at the same time
closed and f\/lat. Dealing with a topologically trivial conf\/iguration
space~$\R^n$, it implies that $\bfA$ is simultaneously exact and pure gauge.
The exactness is already part of the def\/inition~(\ref{matrixdefs}),
and the pure-gauge property is what relates~$\bfA$ with~$\bfu$.
We remark that~$\bfA$ and~$\bff$ are symmetric matrices while $\bfu$ is not.
Furthermore, the inhomogeneity~(\ref{w1}b) demands that $x^i\pa_i\bff=\unity$.
The task is to solve~(\ref{int5}) for $\bff$ and for $\bfu$, which then yield~$\pa^3F$   and   $\vec\nabla U=-2\bfu^{-1}\vec g$.

Of course, we cannot `solve' the WDVV equation by formal manipulations.
But even given a solution~$\bfA$ (and hence $\bff$), it is nontrivial to
construct an associated matrix function~$\bfu$. For this, we must integrate
the linear matrix dif\/ferential equation~(\ref{ulin}),
\begin{gather}\label{bfulin}
\diff\bfu^\top = \bfA \bfu^\top  ,
\end{gather}
which qualif\/ies $\bfu$ as covariantly constant in the WDVV background.
The formal solution reads
\begin{gather*}
\bfu^\top = \sum_{k=0}^\infty \bff^{(k)} \qquad\textrm{with}\qquad
\bff^{(0)}=\unity ,\quad \bff^{(1)}=\bff\quad\textrm{and}\quad
\diff\bff \bff^{(k)}=\diff\bff^{(k+1)}   ,
\end{gather*}
up to right multiplication with a constant matrix.
The matrix functions~$\bff^{(k)}$ are local because
\begin{gather*}
\diff(\diff\bff \bff^{(k)}) = -\diff\bff\wedge\diff\bff^{(k)}=
-\diff\bff\wedge\diff\bff \bff^{(k-1)} = - \bfA\wedge\bfA \bff^{(k-1)} = 0
\end{gather*}
due to the WDVV equation. Likewise, one has
\[
\bff \diff\bff^{(k)} = \bff \diff\bff \bff^{(k-1)} =
\diff\big( \bff \bff^{(k)}-\bff^{(k+1)} \big)  .
\]
Note that the naive guess   $\bfu^\top=\eu^{\bff}$   is wrong since
$[\bff,\diff\bff]=\diff(\bff^2-2\bff^{(2)})\neq0$.

We provide two explicit examples for $n=2$, with the notation
\[
x^{i=1}=:x   ,\qquad x^{i=2}=:y \qquad \mbox{and} \qquad  x^2+y^2=:r^2  .
\]
Starting from the $B_2$ solution with a radial term~\cite{glp2,klp}
\begin{gather} \label{FB2}
F = -\sfrac12 x^2\ln x -\sfrac12 y^2\ln y
-\sfrac14(x + y)^2\ln(x + y) -\sfrac14(x - y)^2\ln(x - y)
+\sfrac12 r^2 \ln r  ,
\end{gather}
we have
\[
\bff=\frac12\left(\begin{matrix}
\ln\bigl[(x^2 - y^2)\sfrac{x^2}{r^2}\bigr] & \ln\sfrac{x+y}{x-y} \\
\ln\sfrac{x+y}{x-y} & \ln\bigl[(x^2 - y^2)\sfrac{y^2}{r^2}\bigr]
\end{matrix}\right)   -  \frac{1}{r^2}\left(\begin{matrix}
x^2 & xy \\[2pt] xy & y^2 \end{matrix}\right)
\]
with   $(x\pa_x+y\pa_y)\bff=\unity$   and, hence,
\[
\bfA=\diff\bff=\frac{(x^2 - y^2)^2}{x y r^4} \left(\begin{matrix}
y\diff x & 0 \\[2pt] 0 & x\diff y \end{matrix}\right)
\ +\ \frac{4 x^2 y^2}{(x^2 - y^2) r^4} \left(\begin{matrix}
x\diff x - y\diff y & x\diff y - y\diff x \\[2pt]
x\diff y - y\diff x & x\diff x - y\diff y \end{matrix}\right)  .
\]
It is easy to check that indeed
$\bfA\wedge\bfA=0$   but   $[\bfA,\bff]\neq0$.
The solution to (\ref{bfulin}) turns out to be
\[
\bfu = \frac{\Gamma}{r^4}\,\left(\begin{matrix}
x r^4 & y r^4 \\[4pt] x\,y^4 & y\,x^4 \end{matrix}\right)
\qquad\buildrel{\Gamma = \unity}\over\longrightarrow\qquad
\begin{cases} u^1=\sfrac12 r^2, \\[2pt]
               u^2=\sfrac12 x^2y^2/r^2 \end{cases}
\]
with an arbitrary non-degenerate constant matrix~$\Gamma$,
as may be checked by inserting it into~(\ref{bfulin}).

One may also begin with a purely radial WDVV solution~\cite{klp},
\[
F = -\sfrac12 r^2 \ln r \qquad\longrightarrow\qquad
\bff= \sfrac12(\ln r^2) \unity  +
\frac{x^2 - y^2}{2 r^2} \sigma_3  +  \frac{xy}{r^2} \sigma_1  ,
\]
and f\/ind
\[
\bfu = \Gamma \left(\begin{matrix}
2x & 2y\\ 2x\arctan\sfrac{y}{x}-y & 2y\arctan\sfrac{y}{x}+x \end{matrix}\right)
\qquad\buildrel{\Gamma = \unity}\over\longrightarrow\qquad
\begin{cases} u^1= r^2, \\[2pt]
               u^2= r^2\arctan\sfrac{y}{x}. \end{cases}
\]
For more generic weight factors in~(\ref{FB2}), $u^2$ is expressed in terms of
hypergeometric functions~\cite{klp}.

\section{Three- and four-particle solutions}
\noindent
An alternative method for constructing solutions $(F,U)$ attempts to
f\/ind functions~$u^A(x)$ satisfying~(\ref{int2}). It is successful
for $n + 1=3$ since the WDVV equation is empty in this case.
Imposing also permutation invariance, a natural choice for three
homogeneous quadratic symmetric functions of $(x^i)=(x,y,z)$ is
\begin{gather}
u^1 =(x + y + z)^2  , \nonumber\\
u^2 =(x - y)^2+(y - z)^2+(z - x)^2 ,\nonumber\\
u^3 =[(2x - y - z)(2y - z - x)(2z - x - y)]^{2/3}  h(s)  ,\label{n3u}
\end{gather}
where $h$ is an (almost) arbitrary function of the ratio
\[
s=\frac{[(2x - y - z)(2y - z - x)(2z - x - y)]^2}
{[(x - y)^2+(y - z)^2+(z - x)^2]^3}  .
\]
Not surprisingly, (\ref{n3u}) fulf\/ils the integrability condition~(\ref{int2}),
so we are guaranteed to produce solutions.
It is straightforward to compute the Jacobians $u^A_{\ i}$ and $w_{A,i}$ and
proceed to the prepotentials. Writing $(g^A)=(g_1,g_2,g_3)$,
the bosonic potential comes out as
\begin{gather*}
V_{\textrm{B}}^{\textrm{cl}}  =  \frac{g_1^2/24}{(x+y+z)^2}
+\frac1{324}\left[ (1-2{s})g_2^2+2{s}
\frac{({h}g_2-g_3/{\root3\of s})^2}{({h}+3{s h'})^2}\right]\\
\phantom{V_{\textrm{B}}^{\textrm{cl}}=}{}
\times
\left(\frac1{(x-y)^2}+\frac1{(y-z)^2}+\frac1{(z-x)^2}\right) \\
 \phantom{V_{\textrm{B}}^{\textrm{cl}}}{}
 =  \frac{g_1^2/24}{(x+y+z)^2}
+\frac{g_2^2-4 s^{\frac23-\de} g_2g_3+2 s^{\frac13-2\de} g_3^2}
{324(1+3\de)^2}
\left(\frac1{(x-y)^2}+\frac1{(y-z)^2}+\frac1{(z-x)^2}\right)\\
\phantom{V_{\textrm{B}}^{\textrm{cl}}=}{}
+\frac{\de(2+3\de)}{8(1+3\de)^2}
\frac{g_2^2}{(x-y)^2+(y-z)^2+(z-x)^2} ,
\end{gather*}
where in the second equality we specialized to
\[
h(s)=s^\de \qquad\longleftrightarrow\qquad
u^3=\frac{[(2x-y-z)(2y-z-x)(2z-x-y)]^{2/3+2\de}}
{[(x-y)^2+(y-z)^2+(z-x)^2]^{3\de}} .
\]
Putting $g_3 = 0$ for simplicity, the corresponding prepotentials are
\begin{gather*}
U = -\frac{g_1}{6}\ln(x + y + z)
-\frac{g_2}{18(1 + 3\de)} \ln(x - y)(y - z)(z - x)\\
\hphantom{U =}{}
-\frac{\de g_2}{4(1 + 3\de)}\ln\big[(x - y)^2 + (y - z)^2 + (z - x)^2\big]  ,
\\
F  = -\tfrac16(x + y + z)^2\ln(x + y + z) - \tfrac14
\bigl[(x - y)^2\ln(x - y)+(y - z)^2\ln(y - z)
\\
\phantom{F=}{}+(z - x)^2\ln(z - x)\bigr] + \tfrac{1 - 6\de}{36}
\bigl[ (2x - y - z)^2\ln(2x - y - z) \\
\phantom{F=}{}
+ (2y - z - x)^2\ln(2y - z - x)
+ (2z - x - y)^2\ln(2z - x - y) \bigr]\\
\phantom{F=}{}
 + \tfrac{\de}{4}
\big[(x - y)^2 + (y - z)^2 + (z - x)^2]\ln[(x - y)^2 + (y - z)^2 + (z - x)^2\big]   .
\end{gather*}
We recognize the roots of $G_2$ plus a radial term
in the coordinate dif\/ferences. The potential simplif\/ies in two special cases:
\begin{alignat*}{3}
& \de=0 \quad \Leftrightarrow\quad h=1   : \qquad &&
V_{\textrm{B}}^{\textrm{cl}}(g_1 = g_3 = 0)\quad\textrm{is pure Calogero}  , & \\
& \de=\sfrac16 \quad \Leftrightarrow\quad h=s^{1/6} : \qquad &&
V_{\textrm{B}}^{\textrm{cl}}(g_1 = g_2 = 0)\quad\textrm{is pure Calogero}  . &
\end{alignat*}
In the full quantum potential,
$V_{\textrm{B}}=V_{\textrm{B}}^{\textrm{cl}}+
\sfrac{\hbar^2}{8}F^{\prime\prime\prime}F^{\prime\prime\prime}$,
the couplings $g^A$ receive quantum corrections.

Stepping up to four particles, i.e.~$n + 1=4$, it becomes much more dif\/f\/icult
to construct solutions, since the integrability condition is no longer trivial.
Our attempts to take a known WDVV solution and exploit the linear equations~(\ref{ulin}) for~$u^A_{\ i}$ have met with success only sporadically. In
most cases, the hypergeometric function~${}_2F_1$ turns up in the expressions.
A simple permutation-symmetric example uses the $A_3$ solution with a radial
term,
\begin{gather*} 
F= -\sfrac18\biggl(\!\sum_i x^i\!\biggr)^2\! \ln \sum_i x^i\!
 + \sfrac18\sum_{i<j} (x^i - x^j)^2 \ln(x^i - x^j) \!
 - \sfrac18\biggl(\!\sum_{i<j}(x^i - x^j)^2\!\biggr)\!
\ln \sum_{i<j} (x^i - x^j)^2  ,
\end{gather*}
for which we discovered~\cite{klp}
\begin{gather*}
u^1  = (x + y + z + w)^2  , \\
u^2  = (x - y)^2+(x - z)^2+(x - w)^2+(y - z)^2+(y - w)^2+(z - w)^2  , \\
u^3  = u^2  I\left(\frac{x+y-z-w}{p q}\right) \qquad \mbox{and} \qquad
u^4  = u^2  I\left(\frac{p}{q}\right)  ,
\end{gather*}
with
 \begin{gather*}
p^2 = (x - y + z - w)+2\sqrt{(w - x)(y - z)},\qquad
q^2 = (x - y - z + w)+2\sqrt{(w - y)(x - z)}
\\
\mbox{and} \qquad I(x)=\int_0^x \frac{\diff t}{\sqrt{1-t^4}} .
\end{gather*}
The Jacobians and the bosonic potential are algebraic but not of Calogero type.
It remains a~challenge to f\/ind $(u^2,u^3,u^4)$ for the $A_3$ WDVV solution
{\it without} radial term,
\[
F= -\sfrac18\biggl( \sum_i x^i \biggr)^2 \ln \sum_i x^i
 - \sfrac18\sum_{i<j} (x^i - x^j)^2 \ln(x^i - x^j)  .
\]

\section[Covector ansatz for prepotential $F$]{Covector ansatz for prepotential $\boldsymbol{F}$}

For the rest of the presentation,
we concentrate on the WDVV equation in $\R^n$,
\[
(\pa_i\pa_k\pa_p F)(\pa_p\pa_l\pa_j F)-
(\pa_i\pa_l\pa_p F)(\pa_p\pa_k\pa_j F)=0
\qquad\textrm{with}\qquad
(x^i\pa_i - 2) F = -\sfrac12 x^ix^i ,
\]
since, together with $U{\equiv}0$, its solutions already produce
genuine $\cN{=} 4$ superconformal mechanics models.
Leaving aside a possible radial term
\begin{gather}\label{w1again}
F_{\textrm{rad}} = -r^2\ln r \qquad\textrm{with}\qquad
r^2:= \sum_i (x^i)^2  ,
\end{gather}
we employ the standard `rank-one' or `covector' ansatz~\cite{wyl}
\[ 
F = -\sfrac12 \sum_{\a}  (\ax)^2 \ln\ax
\]
containing a set $\{\a\}$ of covectors
\[
\a=(\a_1,\a_2,\ldots,\a_n)
\quad\in(\R^n)^* \quad\textrm{or}\quad \in\ic(\R^n)^*
\qquad\longrightarrow\qquad
\a(x)=\ax=\a_ix^i  ,
\]
subject to the normalization
\begin{gather} \label{norm}
\sum_\a  \a_i \a_j = \d_{ij} \qquad\longleftrightarrow\qquad
\sum_\a  \a\otimes\a=\unity
\end{gather}
which takes care of the inhomogeneity in~(\ref{w1again}).
The WDVV equation turns into an algebraic condition
on the set of covectors~\cite{margra,ves,glp2},
\begin{gather} \label{FF}
\sum_{\a,\b}
\frac{\a \cdot \b}{\ax\,\bx}
(\a_i\b_j - \a_j\b_i)(\a_k\b_l - \a_l\b_k)=0
\qquad\textrm{with}\qquad \a \cdot \b=\de^{ij}\a_i\b_j  .
\end{gather}
Apart from the normalization~(\ref{norm}), the covectors are projective,
so we may think of them as a bunch of rays. Let us denote their number
(the cardinality of $\{\a\}$) by $p$. We may assume that no two covectors
are collinear. Since an orthogonal pair of covectors
does not contribute to the double sum, two mutually orthogonal subsets of
covectors decouple in~(\ref{FF}), and it suf\/f\/ices to consider indecomposable
covector sets. In $n = 2$ dimensions, (\ref{norm}) implies~(\ref{FF}),
but already for the lowest nontrivial dimension $n = 3$ only partial results
are known~\cite{wyl,chaves,bgl,glp1,feives1,feives2,glp2}.

\section{Partial isometry formulation of WDVV}

Let us gain a geometric understanding of~(\ref{FF}).
Each of the $\sfrac12p(p - 1)$ pairs $(\a,\b)$ in the double sum spans
some plane $\pi\sim\a{\wedge}\b\in\Lambda^2((\R^n)^*)$,
but not all of these planes need be dif\/ferent. When we group the pairs
according to these planes\footnote{A given covector may occur in dif\/ferent pairs, thus in dif\/ferent groups.
Covectors are not grouped, only their pairs.},
the tensor structure $(\a\wedge\b)^{\otimes2}$
of~(\ref{FF}) tells us that this equation must hold separately for the
subset of coplanar covectors pertaining to each plane~$\pi$,
\begin{gather} \label{WDVVplanar}
\sum_{\a,\b\in\pi}
\frac{\a \cdot \b\ |\a \wedge \b|^2}{\ax \, \bx}=0 \qquad \forall\, \pi  .
\end{gather}
Depending on the number $q$ of covectors contained in a given plane~$\pi$,
one of three cases occurs~\cite{ves,feives2}:
\begin{gather}
\mbox{case (a) \quad   $\pi$ contains zero or one covector
$\quad\longrightarrow\quad$   equation trivial}, \nonumber \\
\mbox{case (b) \quad   $\pi$ contains two covectors, $\pi\sim\a\wedge\b$
$\quad\longrightarrow\quad$   orthogonality \ $\a\cdot\b=0$},\nonumber \\
\mbox{case (c) \quad   $\pi$ contains $q>2$ covectors
$\quad\longrightarrow\quad$   projector condition on $\pi$}:\nonumber\\
 \label{proj}
\sum_{\a\in\pi} \alpha\otimes\alpha= \lap \unity_\pi =: \lap P_\pi
\qquad\textrm{for}\quad \lap\in\R\quad\textrm{and}\quad
P_\pi^2=P_\pi \quad\textrm{with}\quad\textrm{rank}(P_\pi)=2  .
\end{gather}
The latter is the proper covector normalization for the planar subsystem,
which implies the (trivial) WDVV equation on~$\pi$ to hold.
Establishing the projector condition~(\ref{proj}) simultaneously for all planes
is a nontrivial problem, since covectors usually lie in more than one plane,
which imposes conditions linking the planes.

For a more quantitative formulation, we express~(\ref{proj})
in terms of partial isometries. After introducing a counting index
$a=1,\ldots,p$ for the covectors $\{\a\}=\{\a_1,\ldots,\a_p\}$,
we collect their components in an $n \times p$ matrix~$A$. This def\/ines a map
\[
A: \ \R^p\to\R^n \qquad\textrm{given by}\qquad
A=\bigl(\a_{ia}\bigr)^{i=1,\ldots,n}_{a=1,\ldots,p}
\quad\textrm{with}\quad A\,A^\top=\unity_n ,
\]
encoding the total normalization~(\ref{norm}). For each nontrivial plane $\pi$,
we select all $\a_{a_s}\in\pi$, $s=1,\ldots,q$, via
\[
B_\pi: \ \R^p\to\R^q \qquad\textrm{by}\qquad\{\a_a\}\mapsto\{\a_{a_s}\}
\]
and write the combination
\[
\api: \ R^q\to\R^n \qquad\textrm{by}\qquad \api  :=  A B_\pi^\top=
\bigl(\a_{ia_s}\bigr)^{i=1,\ldots,n}_{s=1,\ldots,q}  .
\]
Our projector condition then reads
\begin{gather} \label{pariso}
\api\apt = \lap P_\pi \qquad\longleftrightarrow\qquad
\apt\api = \lap Q_\pi
\end{gather}
with projectors $P_\pi$ on $\R^n$ and $Q_\pi$ on $\R^q$ of rank two
and multipliers~$\lap$, for any nontrivial plane~$\pi$. Therefore,
$A$ is a WDVV solution if\/f $\frac{\api}{\sqrt{\lap}}$
is a rank-2 partial isometry~(\ref{pariso}) for each nontrivial plane $\pi$!
An alternative version of~(\ref{pariso}) is
\[
\api\apt\api = \lap \api .
\]
Note that $A\neq\api B^{\phantom{\top}}_\pi$.
Since the projectors are of rank~2, we may split~$\api$ over~$\R^2$:
\[
\exists \
D_\pi: \ \R^q\to\R^2 \qquad\textrm{and}\qquad C_\pi: \ \R^2\leftarrow\R^n
\qquad\textrm{such that}\qquad
\api=C^\top_\pi D^{\phantom{\top}}_\pi  .
\]
The situation can be visualized in the following noncommutative diagram:
$$
\xymatrix{
&&& *+<1.5ex>[o][F-]{\R^p}
\ar[dll]_{B_\pi} \ar@2[drr]^{A} &&& \\
{\scriptstyle Q_\pi} \ar@/_2pc/[r] &
*+<1.5ex>[o][F-]{\R^q} \ar @/_2pc/ @{-} [l]
\ar[drr]_{D_\pi} \ar@2[rrrr]|-{A_\pi} &&&&
*+<1.5ex>[o][F-]{\R^n} \ar[dll]^{C_\pi}
\ar @/^2pc/ @{-} [r] & {\scriptstyle P_\pi} \ar@/^2pc/[l] \\
&&& *+<1.4ex>[o][F-]{\R^2} &&& }
$$

We illustrate the partial isometry formulation with the simplest nontrivial
example, which occurs at $n = 3$ and $p = 6$, by providing a one-parameter
family of covectors $\{\a,\b,\g,\a',\b',\g'\}(t)$ via
\begin{gather} \label{parisoex}
A= \frac16 \bordermatrix{
& \a   & \b         & \g         & \a'        & \b'       & \g'       \cr
& 6t   &-3t         &-3t         & 0          & 3w        &-3w        \cr
& 0    & 3\sqrt{3}t &-3\sqrt{3}t &-2\sqrt{3}w & \sqrt{3}w & \sqrt{3}w \cr
& 0    & 0          & 0          & 2\sqrt{3}  & 2\sqrt{3} & 2\sqrt{3} \cr
}
\qquad\textrm{with}\quad w=\sqrt{2-3t^2}  .
\end{gather}
It is easily checked that   $A\,A^\top=\unity$.
A quick analysis of linear dependence reveals that 12 of the 15 covector pairs
are grouped into 4~planes of 3~pairs each, leaving 3 pairs ungrouped.
3~coplanar pairs imply 3~coplanar covectors, hence there are 4 nontrivial
planes containing $q = 3$ covectors, namely
\[
\<\a\,\b\,\g\> ,\qquad \<\a\,\b'\g'\> ,\qquad
\<\a'\b\,\g'\> ,\qquad \<\a'\b'\g\> ,
\]
and 3 planes containing just two covectors, which are indeed orthogonal,
\[
\a\cdot\a^\prime= \b\cdot\b^\prime= \g\cdot\g^\prime= 0 .
\]
Let us test the projector condition~(\ref{pariso}) for two of the planes:
\begin{gather*}
A_{\<\a\,\b\,\g\>}\,=\frac12 \left( \begin{matrix}
2t & -t & -t \\ 0 & \sqrt{3}\,t & \!-\sqrt{3}\,t\! \\ 0 & 0 & 0
\end{matrix} \right) \quad\Rightarrow\qquad
{\api\apt}= {\sfrac32 t^2} \cdot\left( \begin{matrix}
1 & 0 & 0 \\ 0 & 1 & 0 \\ 0 & 0 & 0
\end{matrix} \right) = {\sfrac32 t^2}\cdot{P_\pi}   ,
\\
A_{\<\a\,\b'\g'\>}=\frac16 \left( \begin{matrix}
6t & 3w & -3w \\ 0 & \sqrt{3}w & \sqrt{3}w \\ 0 & 2\sqrt{3} & 2\sqrt{3}
\end{matrix} \right) \quad\Rightarrow\qquad
{\api\apt}= \frac{1 - \sfrac12t^2}{6 - 3t^2} \left( \begin{matrix}
{\scriptstyle 6 - 3t^2} & 0&0 \\ 0 & {\scriptstyle 2 - 3t^2} & 2w \\ 0 & 2w & 4
\end{matrix} \right)  ,
\end{gather*}
where the matrices on the right are idempotent. Hence, in both cases, $\api$
is proportional to a~partial isometry, with a (parameter-dependent) multiplier~$\lap$. The other two nontrivial planes work in the same way. We have proven that~(\ref{parisoex}) produces a family of WDVV solutions. This scheme naturally extends to include imaginary covectors as well.

\section{Deformed root systems and polytopes}

It is known for some time~\cite{margra,ves} that the set $\Phi^+$ of
positive roots of any simple Lie algebra (in fact, of any Coxeter system)
is a good choice for the covectors. So let us take
\[
\{\a\} = \Phi^\+ = \Phi^+_{\textrm{L}}\cup\Phi^+_{\textrm{S}} \qquad
\textrm{with}\qquad \a_{\textrm{L}} \cdot \a_{\textrm{L}}=2 \quad\textrm{and}
\quad \a_{\textrm{S}} \cdot \a_{\textrm{S}}=1\ \textrm{or}\ \sfrac23  .
\]
where the subscripts `L' and `S' pertain to long and short roots, respectively.
Having f\/ixed the root lengths, we must introduce scaling factors
$\{f_\a\} = \{f_{\textrm{L}},f_{\textrm{S}}\}$   in
\begin{gather*} 
F =  -\sfrac12  \left(
f_{\textrm{L}}  \sum_{\a\in\smash{\Phi^+_{\textrm{L}}}} +
f_{\textrm{S}}  \sum_{\a\in\smash{\Phi^+_{\textrm{S}}}} \right)
(\ax)^2 \ln|\ax|  .
\end{gather*}
The normalization condition~(\ref{norm}) has a one-parameter solution,
\begin{gather*}
\left(
f_{\textrm{L}}  \sum_{\a\in\smash{\Phi^+_{\textrm{L}}}} +
f_{\textrm{S}}  \sum_{\a\in\smash{\Phi^+_{\textrm{S}}}} \right)
\a\otimes\a =\unity \qquad\longrightarrow\qquad
\begin{cases}
f_{\textrm{L}}=\dfrac{1}{h^\vee} + (h -  h^\vee) t , \vspace{1mm}\\
f_{\textrm{S}}=\dfrac{1}{h^\vee} + (h - rh^\vee) t,
\end{cases} \quad\textrm{with}\quad t\in\R  ,
\end{gather*}
where $h$ and $h^\vee$ are the Coxeter and dual Coxeter numbers of the
Lie algebra, respectively.
The roots def\/ine a family of over-complete partitions of unity.
Amazingly, all simple Lie algebra root systems obey~(\ref{WDVVplanar}),
and they do so separately for the pairs of long roots, for the pairs of
short roots and for the mixed pairs, of any plane~$\pi$.
This leads to the freedom~($t$) to rescale the short versus the long roots
and provides a one-parameter family of solutions to the WDVV
equation~\cite{ves,strachan,glp2}.
(In the simply-laced case there is only one solution, of course.)

For illustration we give two examples.
Let $\{e_i\}$ be an orthonormal basis in $\R^{n+1}$. For
\begin{gather*}
A_n\oplus A_1 : \quad
\{\a\} = \biggl\{\ e_i - e_j  ,\  \sum_i e_i\ \Big| \
1\le i<j\le n + 1\ \biggr\}
\qquad\textrm{we f\/ind} \\
F_{A_n\oplus A_1}=
-\frac{1/2}{n+1} \sum_{i<j} (x^i - x^j)^2 \ln(x^i - x^j)\ -\
\frac{1/2}{n+1}\bigg(\sum_i x^i\bigg)^2\ln\bigg(\sum_i x^i\bigg)
\end{gather*}
with center-of-mass decoupling,
while for the non-simply-laced case ($n = 2$, $p = 6$) without center of mass
\begin{gather*}
G_2 : \
\{\a\} = \left\{ \sfrac{1}{\sqrt{3}}(e_i - e_{j})  ,\
\sfrac{1}{\sqrt{3}}(e_i + e_{j} - 2e_{k})
\ \Big| \  (i,j,k)\quad\textrm{cyclic}  \right\}
\qquad\textrm{one gets} \\
F_{G_2}=-\frac{1-24t}{24} \big(x^1 - x^2\big)^2 \ln\big(x^1 - x^2\big)
  -  \frac{1+8t}{24}\big(x^1 + x^2 - 2x^3\big)^2 \ln\big(2x^1 - x^2 - x^3\big)
\\
\hphantom{F_{G_2}=}{}
  +  \textrm{cyclic}  .
\end{gather*}

A natural question is whether one can deform the Lie algebraic root systems
by changing the angles between covectors but keep~(\ref{WDVVplanar}) valid.
So which deformations respect the WDVV equation? Based on a few examples,
we conjecture that the (suitably rescaled and translated) covectors
should form the edges of some polytope in~$\R^n$. Non-concurrent pairs of
edges then have no reason to be coplanar with other edges, thus better be
orthogonal. Concurrent edge pairs, on the other hand, always belong to some
polytope face, hence automatically combine with further coplanar edges to
a nontrivial plane~$\pi$. The hope is that the polytope's incidence relations
take care of the WDVV equation, e.g.~in the form of~(\ref{pariso}).
For $p\ge\sfrac12n(n + 1)$, there is enough scaling freedom to f\/inally arrange
the normalization~(\ref{norm}) with~$\{f_\a\}$.

\begin{wrapfigure}[6]{l}{3cm}
\vspace{-3mm}
\includegraphics[width=3cm]{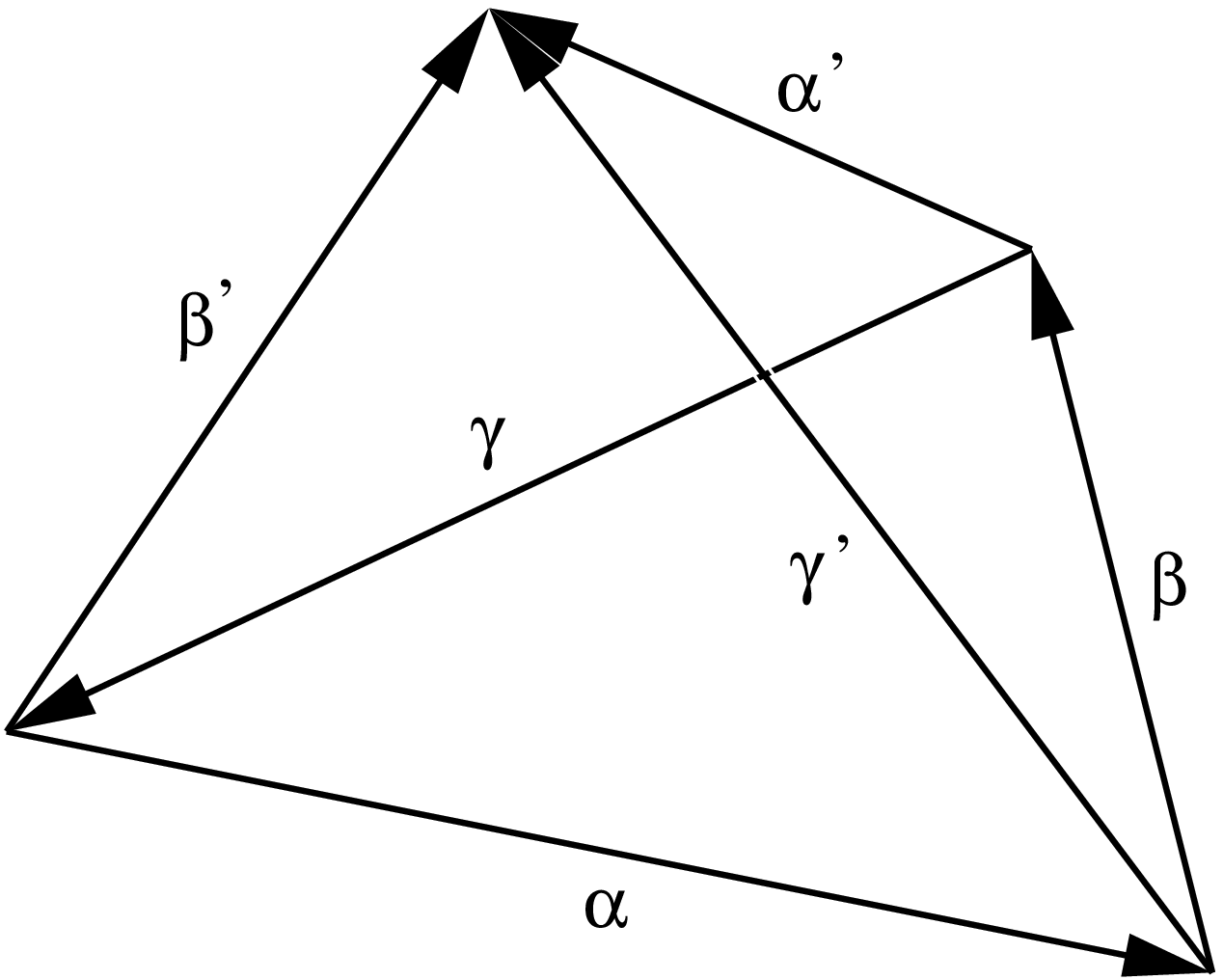}
\end{wrapfigure}
This expectation is actually bourne out in the case of the $A_n$ root system,
which, with $p=\sfrac12n(n + 1)$, is in fact the minimal irreducible system
in each dimension~$n$ and uniquely f\/ixes~$\{f_\a\}$.
Starting with an arbitrary bunch of $\sfrac12n(n + 1)$ rays in~$\R^n$,
we reduce the freedom in their directions by imposing f\/irstly the $n$-simplex
incidence relations and secondly the orthogonality conditions for skew edges.
Let us do some counting of moduli (minus global translations, rotations
and scaling):
\begin{center}
\begin{tabular}{|c|c@{\,\,\,}c@{\,\,\,}c@{\,\,\,}c@{\,\,\,}c|}
\hline
& ray moduli & incidences & simplex moduli & orthogonality & f\/inal moduli \\
\hline
$\#$ & $\sfrac12n^2(n - 1)$ & $-\sfrac12(n - 2)(n^2 - 1)$ &
$\sfrac12(n - 1)(n + 2)$ & $-\sfrac12(n - 2)(n + 1)$ & $n$ \phantom{\Big|} \\
\hline
$\!\!n = 2,3,4\!\!$ &
$2,\,9,\,24$ & $0,-4,-15$ & $2,\,5,\,9$ & $0,-2,-5$ & $2,\,3,\,4$ \\
\hline
\end{tabular}
\end{center}
We f\/ind that the moduli space ${\cal M}(A_n)$ of these so-called
{\it orthocentric\/} $n$-simplices is just $n$-dimensional.
It can be shown~\cite{letalk} that indeed it f\/its perfectly to a family of
WDVV solutions found earlier~\cite{chaves,feives1,feives2},
lending support to our polytope idea.
We remark that the previous example~(\ref{parisoex})
represents a one-parameter subset in~${\cal M}(A_3)$.

Let us make this observation more explicit in the case of $n = 4$.
Using the recursive construction of orthocentric $n$-simplices presented
in~\cite{glp2} for $n = 4$ and computing the corresponding scaling factors
is feasible but algebraically involved. Therefore, we just present a `nice'
one-parameter subfamily of solutions, with $t\in\R_+$ and $w^2=t^2-\sfrac14$,
\begin{gather*}
A = \frac{1}{2t} \!\left(\!\begin{matrix}
           w\sqrt{2} &  0                   &  \frac{w}{2}\sqrt{2} &
-\frac{w}{2}\sqrt{2} & -\frac{w}{2}\sqrt{2} &  \frac{w}{2}\sqrt{2} &
\ph\frac12\sqrt{2}& -\frac12\sqrt{2} &  0               &  0              \\[4pt]
 0                 &\!\!-\frac{w}{3}\sqrt{6}&  \frac{w}{2}\sqrt{6} &
\ph\frac{w}{6}\sqrt{6}&\ph\frac{w}{2}\sqrt{6}& \frac{w}{6}\sqrt{6} &
\ph\frac16\sqrt{6}&\ph\frac16\sqrt{6}& -\frac13\sqrt{6} &  0              \\[4pt]
 0                   & \frac{2w}{3}\sqrt{3} &  0                   &
\frac{2w}{3}\sqrt{3} &  0                   & \frac{2w}{3}\sqrt{3} &
\ph\frac16\sqrt{3}&\ph\frac16\sqrt{3}&\ph\frac16\sqrt{3}& -\frac12\sqrt{3}\\[4pt]
0 & 0 & 0 & 0 & 0 & 0 & \ph t &\ph t &\ph t & t \end{matrix}\!\right)\!.
\end{gather*}
For $t^2 = \sfrac54$ we have the root system of~$A_4$, at $t^2 = \sfrac14$
the f\/irst six covectors disappear and leave~$A_1^4$. When $0<t^2<\sfrac14$,
the f\/irst six covectors are imaginary, and in the singular limit $t^2{\to}0$
we obtain the $A_3$ roots and fundamental weights, but can no longer maintain
our normalization.

A more familiar parametrization embeds the $A_4$ root system into $\R^5$,
in the hyperplane orthogonal to the center-of-mass covector~$\sum_i e_i$,
with $s\in\R_+$ and $u^2=20s^2 - 10s + 1$,
\[
A = \frac{1}{(1 - 4s)\sqrt{5}} \left( \begin{matrix}
 u & 0 & u & 0 & u & 0 & \ph 1 - s & -s & -s & -s \\[2pt]
 -u & 0 & 0 & u & 0 & u    & \ph -s & 1 - s & -s & -s \\[2pt]
0 & u &  -u & 0 & 0 &  -u       & \ph -s & -s & 1 - s & -s \\[2pt]
0 &  -u & 0 &  -u &  -u & 0          & \ph -s & -s & -s & 1 - s \\[2pt]
0 & 0 & 0 & 0 & 0 & 0             & \ph 4s - 1&4s - 1&4s - 1&4s - 1
\end{matrix} \right) .
\]
Now $s = 0$ yields the roots of $A_4$, beyond
$s=\sfrac14(1 - \sfrac1{\sqrt{5}})$ the f\/irst six covectors turn imaginary,
and the singular limit $s \to \sfrac14$ ($u \to \sfrac{\ic}2$) gives
the $A_3$ roots and fundamental weights, orthogonal also to $\sum_i e_i-5e_5$.
This pattern generalizes to an interpolation between the $A_n$ roots and the
$A_{n-1}$ roots and fundamental weights.

\begin{wrapfigure}[5]{r}{3cm}
\vspace{-7mm}
\includegraphics[width=3cm]{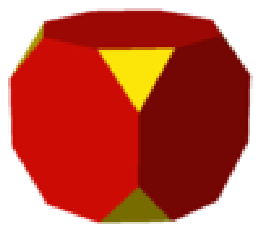}
\end{wrapfigure}
What about deformations of other root or weight systems?
We give two more prominent examples in $n = 3$ dimensions. First,
consider the $p = 9$ positive roots of $B_3$ and observe that,
from four copies of them, we can assemble the edges of a truncated cube.
It is possible to deform the latter into a truncated cuboid while
keeping the orthogonalities and producing a~six-parameter family of covectors,
\begin{gather*}
\bigl\{\ax\bigr\} = \bigr\{
d_1x^1,\; d_2x^2,\;  d_3x^3 ;\;
c_3(c_2x^1{\pm}c_1x^2) ,\;  c_1(c_3x^2{\pm}c_2x^3) ,\; c_2(c_1x^3{\pm}c_3x^1)
\bigr\} ,\quad c_i,d_i\in\R  .
\end{gather*}
The normalization   $\sum_\a f_\a \a{\otimes}\a=\unity$   can be achieved with
\begin{gather*}
\bigl\{ f_\a \bigr\} = \left\{
\frac{c_0^2+c_1^2-c_2^2-c_3^2}{c^2\ d_1^2} ,
\frac{c_0^2-c_1^2+c_2^2-c_3^2}{c^2\ d_1^2} ,
\frac{c_0^2-c_1^2-c_2^2+c_3^2}{c^2\ d_1^2} ;
\frac{1}{c^2\,c_3^2} , \frac{1}{c^2\,c_1^2} , \frac{1}{c^2\,c_2^2}
\right\} ,\\
c^2=c_0^2 + c_1^2 + c_2^2 + c_3^2  .
\end{gather*}
One sees that the relevant combinations $\sqrt{f_\a} \a$ depend only
the three ratios $\sfrac{c_i}{c_0}$. It turns out that we have
constructed a three-dimensional moduli space of WDVV solutions~\cite{chaves,feives1,feives2}.

\begin{wrapfigure}[5]{r}{3cm}
\vspace{-5mm}
\includegraphics[width=3cm]{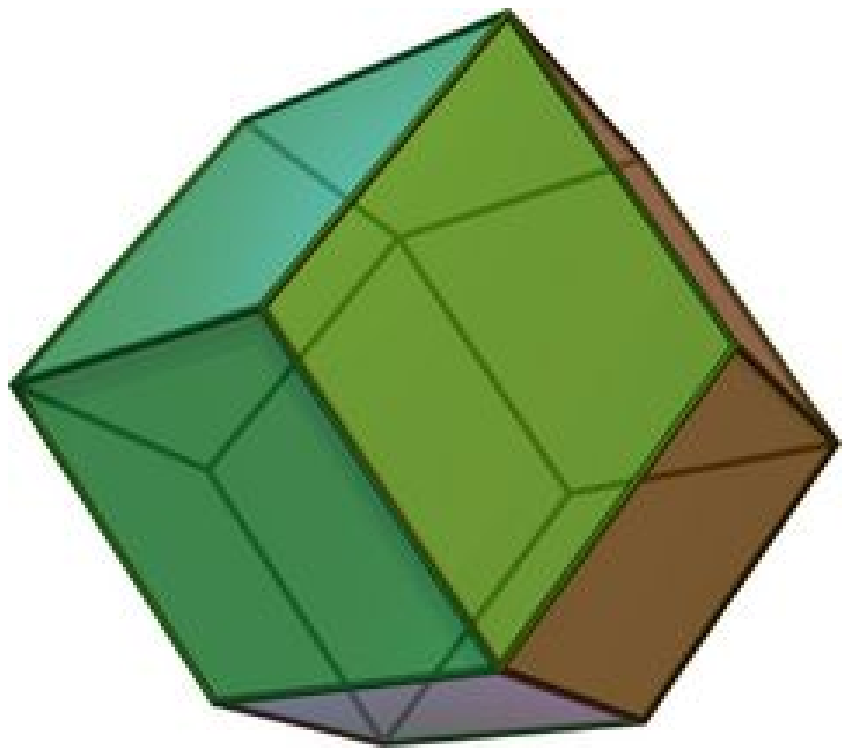}
\end{wrapfigure}
Second, again using $A_3$, it is possible to combine four copies of its
three positive vector weights with six copies of its four positive spinor
weights to the edge set of a rhombic dodecahedron, with each rhombic face
being dissected into two triangles. There exists a three-parameter family
of deformations in line with the orthogonalities, given by
\begin{gather*}
\ax=d_1 x^1,\qquad \bx=d_2 x^2,\qquad \gx=d_3 x^3 ;\\
\frac{\a + \b + \g}{2}  ,\qquad \frac{\a - \b - \g}{2}  ,\qquad
\frac{-\a + \b - \g}{2}  ,\qquad \frac{-\a - \b + \g}{2}  ,
\end{gather*}
and re-normalization is achieved by
\begin{gather*}
f_\a=\frac{-d_1^2+d_2^2+d_3^2}{d^2 d_1^2},\qquad
f_\b=\frac{ d_1^2-d_2^2+d_3^2}{d^2 d_2^2},\qquad
f_\g=\frac{ d_1^2+d_2^2-d_3^2}{d^2 d_3^2};\\
f_{\textrm{spinor}}= \frac{2}{d^2} ,\qquad
d^2=d_1^2 + d_2^2 + d_3^2 .
\end{gather*}
In this case, the combinations $\sqrt{f_\a} \a$
depend only on the ratios $\sfrac{d_i}{d_j}$, and we again discover
a~two-dimensional family of WDVV solutions~\cite{feives1,feives2}.
It seems that indeed the polytope's incidence relations imply the WDVV
equation, thus allowing us to construct solutions~$F$ purely geometrically,
by guessing appropriate polytopes with certain edge multiplicities.

\section{Hypergraphs}

Sadly, our ortho-polytope concept fails, as may be seen from the
f\/irst counterexample at $(n,p)=(3,10)$:
\begin{gather} \label{p10ex}
A = \frac1{4\sqrt{3}} \bordermatrix{
& 1 & 2 & 3 & 4 & 5 & 6 & 7 & 8 & 9 & 10             \cr
& 2\sqrt{3} & \ph2\sqrt{3} & 2\sqrt{2}   & \ph0\ph & \sqrt{2} & -\sqrt{2}   &
   \sqrt{6} &  -\sqrt{6}   & \ph0        & 0         \cr
& 2\sqrt{2} & -2\sqrt{2}   & 0           & 4       & \sqrt{3} & \ph\sqrt{3} &
 -1         & -1           & -\sqrt{6}   &  \sqrt{2} \cr
& 0         &  0           & 0           & 0       & \sqrt{3} & \ph\sqrt{3} &
 \ph3       & \ph3         & \ph\sqrt{6} & 3\sqrt{2}
}
\end{gather}
\begin{wrapfigure}[7]{l}{6.1cm}
\vspace{-6mm}
\includegraphics[width=5.8cm]{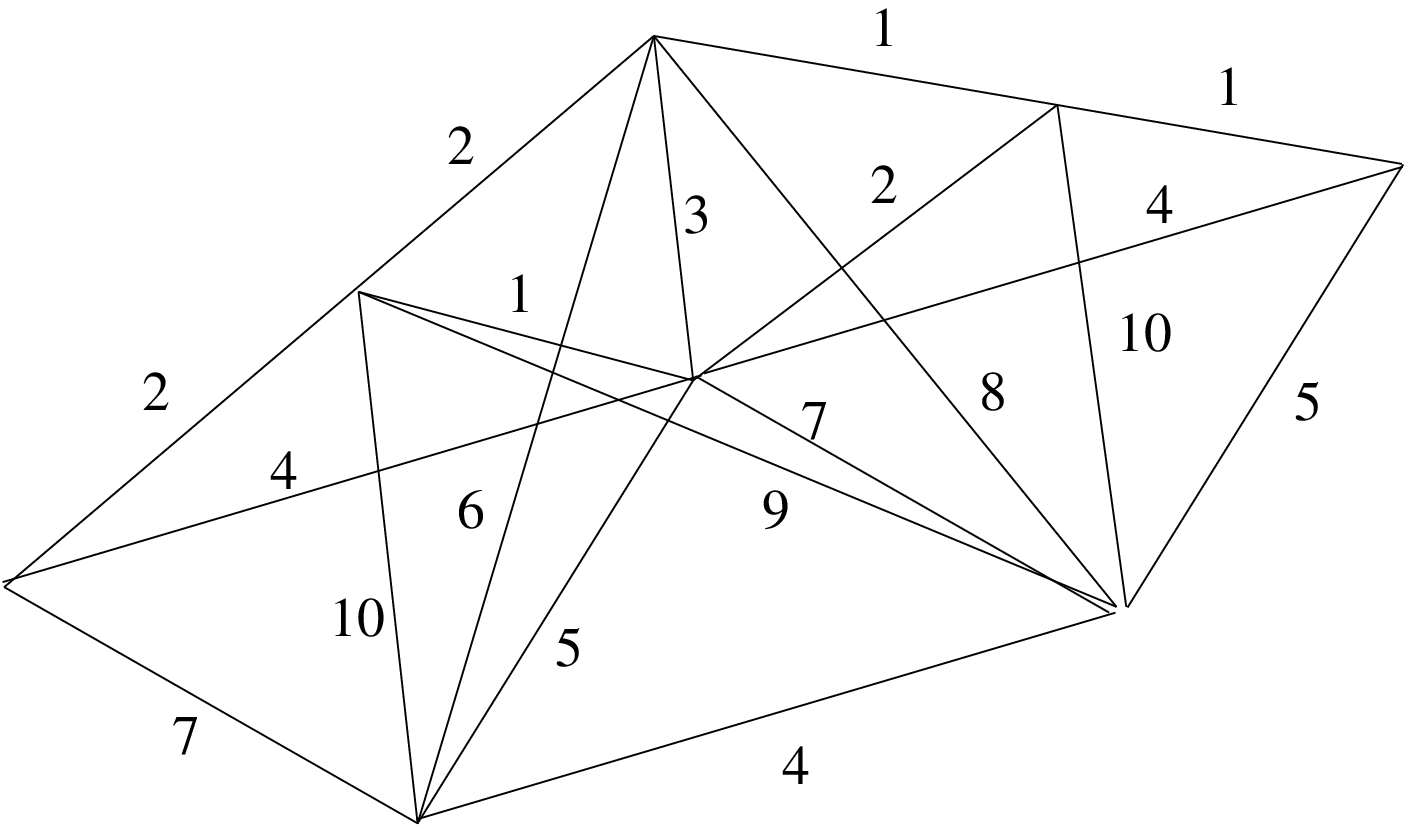}
\end{wrapfigure}
is properly normalized, $A A^\top=\unity_3$, and may be checked to fulf\/il
the partial-isometry conditions~(\ref{pariso}) for each nontrivial plane.
It turns out, however, that there exists no polyhedron whose edges are built
from (suitably rescaled copies of) all ten column vectors in~(\ref{p10ex}).
In the attempt shown to the left, one of the would-be edges (labelled `9')
runs {\it inside\/} the convex hull created by the others.

This lesson demonstrates that it may be better to restrict ourselves to
the essential feature of~$A$, which is the coplanarity property of its
columns~$\a$. Even though there is not always an ortho-polytope, we may still
hope that each $n \times p$ matrix~$A$ with
\[
A A^\top=\unity_n \qquad \mbox{and} \qquad
q(\a_a{\wedge}\a_b)=2 \quad\Rightarrow\quad
\a_a\cdot\a_b=0 \qquad\forall\, a,b=1,\ldots,p
\]
already obeys the crucial conditions~(\ref{pariso}) for all $q > 2$ planes.

\begin{wrapfigure}[9]{l}{4cm}
\vspace{-3mm}
\includegraphics[width=4cm]{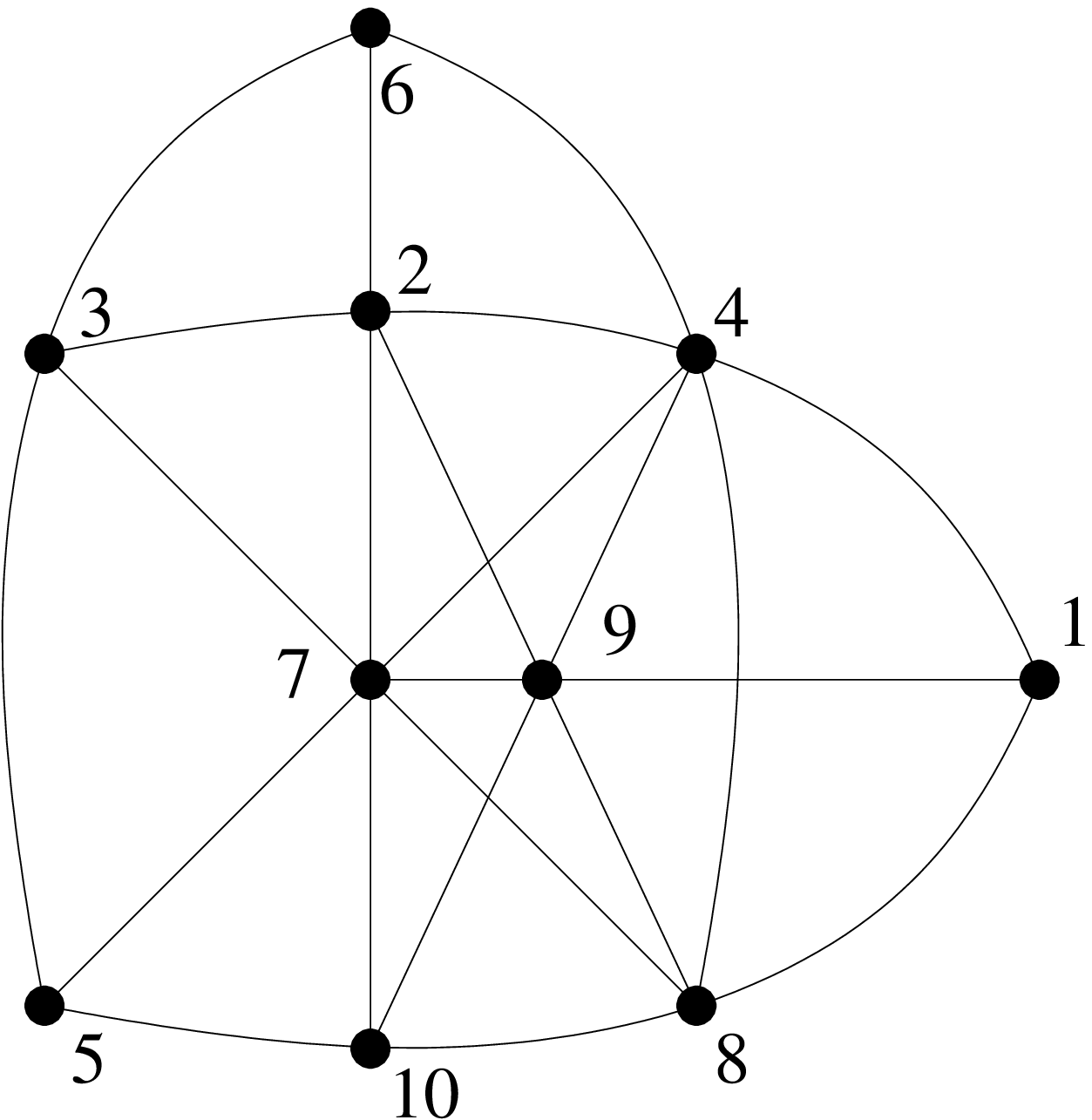}
\end{wrapfigure}Suppose we have $m$ nontrivial planes and label them by $\mu=1,\ldots,m$.
The $q_\mu > 2$ covectors in the plane~$\pi_\mu$ are grouped in the subset
$\{\a_{a_s^\mu}\}\subset\{\a_a\}$, with $s=1,\ldots,q_\mu$.
A shorter way of encoding this coplanarity information
is by using only the {\it labels\/} rather than denoting the covectors.
Thus, we combine the $a_s^\mu$ for each nontrivial plane~$\pi_\mu$ in the
subset $\{a_s^\mu|s = 1,\ldots,q_\mu\}=:\Pi_\mu\subset\{1,\ldots,p\}$,
and then write down the collection
$H(A):=\{\Pi_1,\Pi_2,\ldots,\Pi_m\}\subset{\cal P}(\{1,\ldots,p\})$
of these (overlapping) subsets.
Such subset collections are known as {\it simple hypergraphs\/}~\cite{hyper}.
They are graphically represented by \mbox{writing} a vertex for each covector label
and then, for each~$\mu$, by connecting all vertices whose labels occur in~$\Pi_\mu$. The resulting graph has $p$~vertices and $m$~connections~$\Pi_\mu$,
called {\it hyperedges}. Note that a vertex represents a covector, and a
hyperedge stands for a (nontrivial) plane, thus gaining us one dimension in
drawing\footnote{Our simple hypergraphs contain only $q_\mu$-vertex hyperedges with
$q_\mu>2$, hence no one- or two-vertex hyperedges.}.
As an example, the hypergraph for~(\ref{p10ex})~reads
$\{\{1234\}\{1580\}\{2670\}\{179\}\{289\}\{356\}\{378\}\{457\}\{468\}\{490\}\}$
and is represented above (with `$0$'=`$10$').
To the mathematically inclined reader, we note that our simple hypergraphs are
not of the most general kind: they are also
\begin{itemize}\itemsep=0pt
\item {\it linear\/}:
the intersection of two hyperedges has at most one vertex
(uniqueness of planes)
\item {\it irreducible\/}:
the hypergraph is connected
(the covector set does not decompose)
\item {\it complete\/}:
when adding the $q_\mu  =  2$ planes,
each vertex pair is contained in a~hyperedge
\item {\it orthogonal\/}:
a nonconnected vertex pair is `orthogonal' (property of the $q_\mu = 2$ planes)
\end{itemize}
Of course, two hypergraphs related by a permutation of labels are equivalent.
Thus, our program is to construct, for a given value of~$p$, all
orthogonal complete irreducible linear simple hypergraphs and check the
partial-isometry conditions~(\ref{pariso}) for each plane~$\pi$.
Unfortunately, this is not so easy, because
the orthogonality is not a natural hypergraph property but depends on
the dimension~$n$ of a~possible covector realization.
In fact, it is not guaranteed that such a~realization exists at all.
Therefore, the classif\/ication of complete irreducible linear simple hypergraphs
with $p$~vertices has to be amended by the construction of the corresponding
covector sets in~$\R^n$, subject to the orthogonality condition.

\section{Matroids}

Luckily, there is another mathematical concept which abstractly captures the
linear dependence in a subset of a power set, namely the notion of a {\it matroid}
\cite{matroid,ox,dukes}.
There exist several equivalent def\/initions of a matroid, for example as the
collection $\{C_\mu\}$ of all {\it circuits\/} $C_\mu\subset\{1,\ldots,p\}$,
which are the minimal dependent subsets of our ground set~$\{1,\ldots,p\}$:
\begin{itemize}\itemsep=0pt
\item The empty set is not a circuit.
\item No circuit is contained in another circuit.
\item If $C_1\neq C_2$ share an element~$e$,
then $(C_1\cup C_2)\backslash\{e\}$ is or contains another circuit.
\end{itemize}
Of course, we identify matroids related by permutations of the ground set.

The idea is that each circuit corresponds to a subset of linearly dependent
covectors. Indeed, every $n \times p$ matrix~$A$ produces a matroid.
However, the converse is false: not every matroid is representable
in some~$\R^n$. If so, it is called an {\it $\R$-vector\/} matroid,
with rank $r\le n$.
The rank $r_\mu=|C_\mu|-1$ of an individual circuit~$C_\mu$ is the dimension
of the vector space spanned by its covectors. Excluding one- and
two-element circuits qualif\/ies our matroids as {\it simple\/}.
It may happen that two rank-$d$ circuits span the same vector space,
for example if they agree in $d$ of their elements.
Hence, it is useful to unite all rank-$d$ circuits spanning the same
$d$-dimensional subspace in a so-called {\it $d$-flat\/}~$F_d$,
with $2\le d<r$. We call such a $d$-f\/lat {\it minimal\/} if it arises
from a single circuit, i.e.\ $|F_d|=d + 1$.
In this way, we may label the matroid more ef\/f\/iciently by listing all
2-f\/lats, 3-f\/lats etc., all the way up to $r - 1$.
Needless to say, we are only interested in {\it connected\/} matroids,
i.e.\ those which do not decompose as a direct sum.
Also, for a given dimension~$n$ we study only $\R$-vector matroids of rank
$r = n$ and ignore those of smaller rank, since they can already be
represented in a smaller vector space.
Finally, we need to implement the orthogonality property. So let us call
a matroid {\it orthogonal\/}, if any pair of covectors which does not share
a 2-f\/lat is orthogonal. Note that further orthogonalities (inside 2-f\/lats)
may be enforced by the representation.

A matroid of rank~$r$ can be represented geometrically in $\R^{r-1}$ as
follows. Mark a node for every element of the ground set (the covectors).
Then, connect by a line all covectors in one 2-f\/lat, for all 2-f\/lats. Next,
draw a two-surface containing all covectors in one 3-f\/lat, for all 3-f\/lats,
and so on. We illustrate this method on two examples, the $A_4$ and the
$B_3$ matroid:

\centerline{\includegraphics[width=7.5cm]{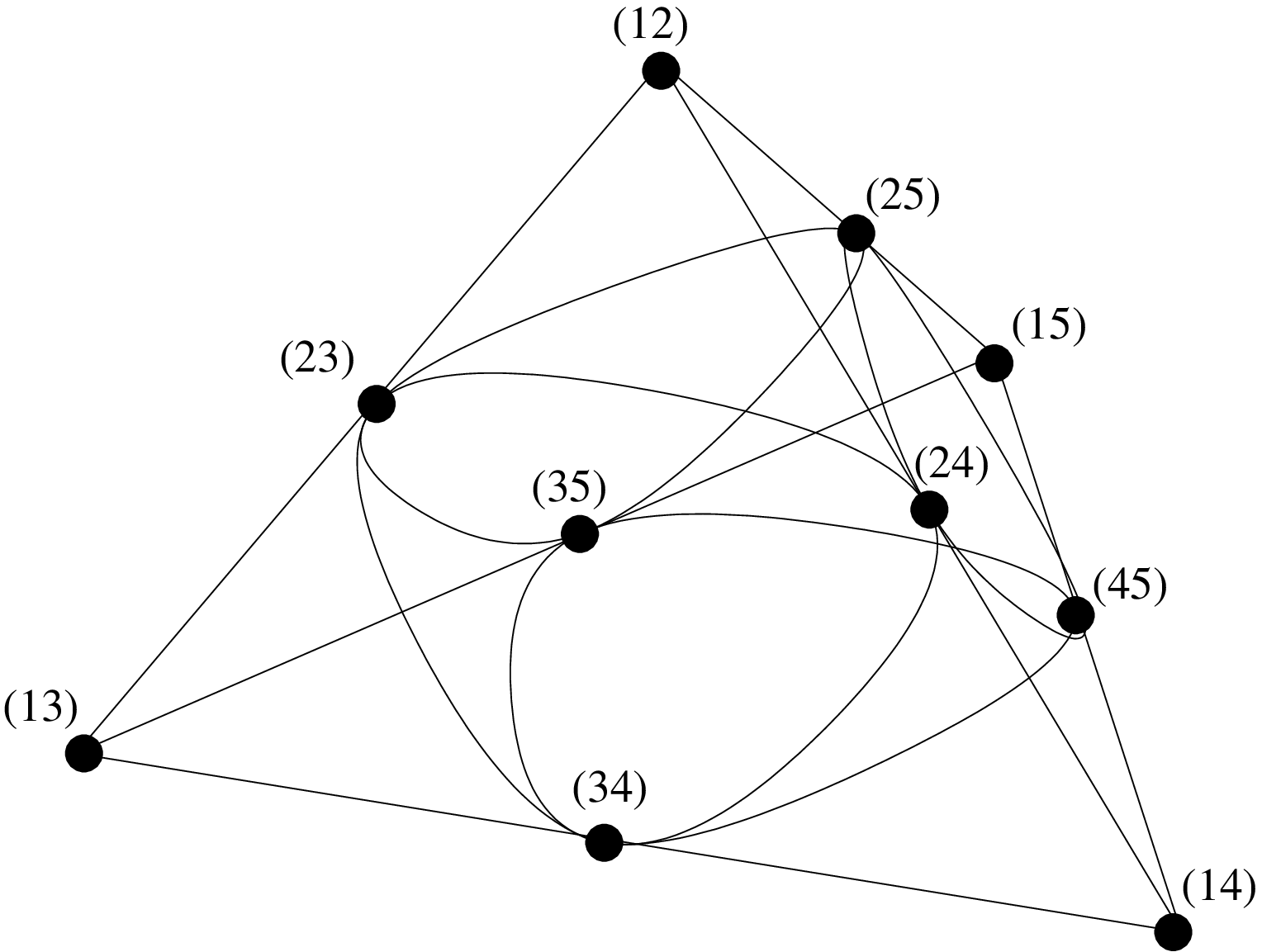}
\qquad
\includegraphics[width=6cm]{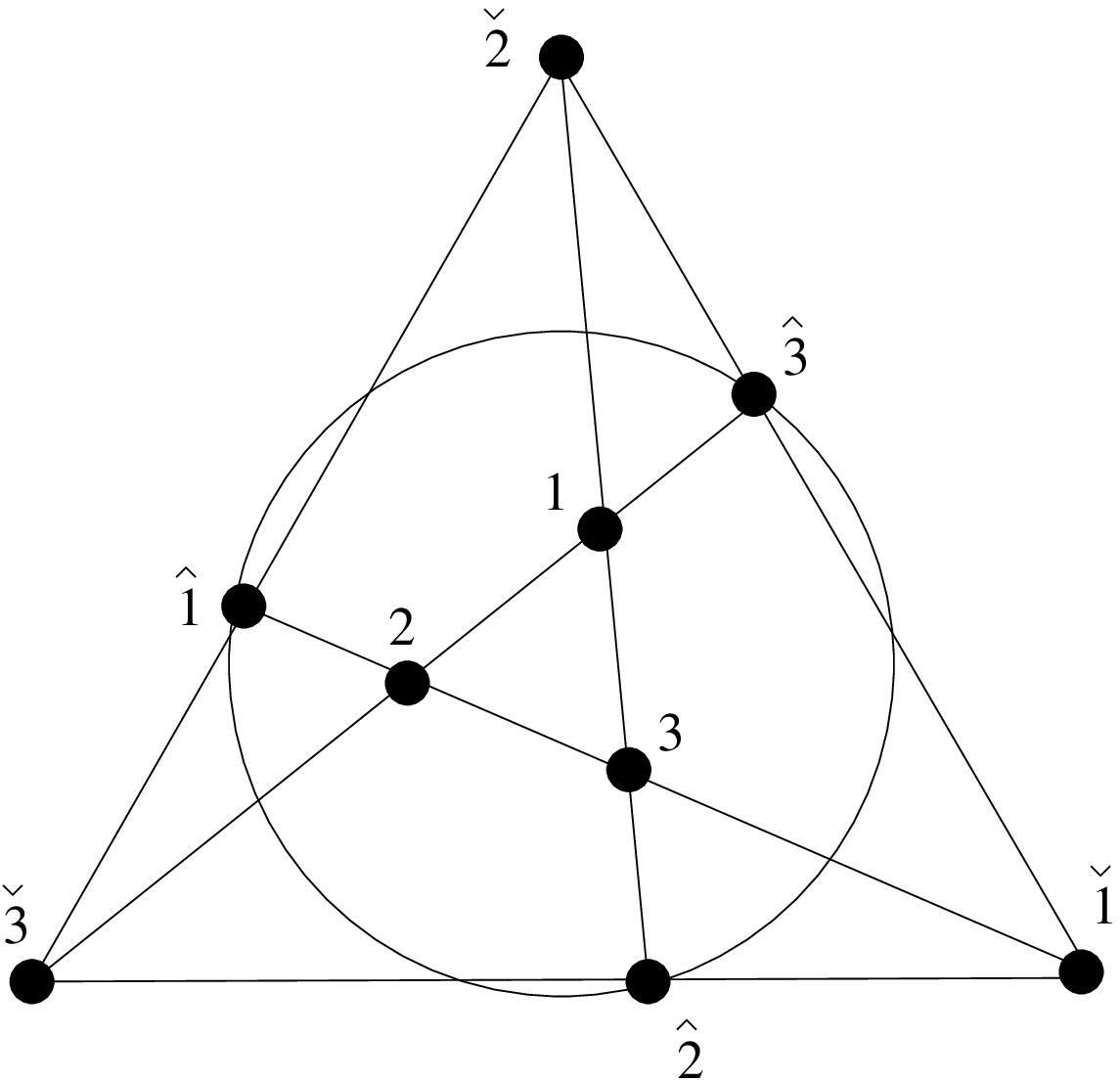}}

The $A_4$ case has $r = 4$, and it is natural to label
the ten covectors by pairs $(ij)$, with $1\le i<j\le5$. Then,
\begin{gather*}
\{C_\mu\}=\{   \{(ij)(ik)(jk)\},
\{(ij)(ik)(j\ell)(k\ell)\},\{(ij)(i\ell)(jk)(k\ell)\},
\{(ik)(i\ell)(jk)(j\ell)\}, \\
\phantom{\{C_\mu\}=\{}{}
\{(1i){\textstyle{i\choose j}{j\choose k}{k\choose\ell}}(1\ell)\}\}
\qquad\textrm{with}\quad {\textstyle{i\choose j}}=(ij)\ \textrm{or}\ (j\,i)
\end{gather*}
lists ten circuits of rank~2, f\/ifteen circuits of rank~3 and twelve circuits
of rank~4. The former represent ten 2-f\/lats, the middle unite in triples to
f\/ive 3-f\/lats and the latter combine to the trivial 4-f\/lat,
\begin{gather*}
\{F_2\} =\{\{(ij)(ik)(jk)\}\}   ,\\
\{F_3\} =\{\{(ij)(ik)(i\ell)(jk)(j\ell)(k\ell)\}\}   ,\\
\{F_4\} =\{\{(12)(13)(14)(15)(23)(24)(25)(34)(35)(45)\}\}  .
\end{gather*}
Orthogonality is required between pairs with fully distinct labels.
The $B_3$ example is of rank three but less symmetric.
We label the three short roots by $i$ and the six long ones by~$\hat{i}$ and~$\check{i}$, with $i=1,2,3$, and obtain sixteen rank-2 circuits grouping
into seven 2-f\/lats and thirty-nine rank-3 circuits combining into the
unique 3-f\/lat ($i\neq j\neq k\neq i$),
\begin{gather*}
\{C_\mu\}=\{
\{ij\hat k\},\{ij\check k\},\{i\hat j\check j\},
\{\hat i\hat j\hat k\},\{\hat i\check j\check k\}, \\
\phantom{\{C_\mu\}=\{  }{}
\{ij\,\hat i\hat j\},\{ij\,\check i\check j\},\{ij\,\hat i\check j\},
\{i\,\hat i\,\check i\hat j\},\{i\,\hat i\,\check i\check j\},
\{i\,\hat i\hat j\check k\},\{i\,\check i\hat j\hat k\},
\{i\,\check i\check j\check k\},\{\hat i\,\check i\hat j\check j\}\}   ,
\\
\{F_2\} =\{
\{1,2,\hat3,\check3\},\{1,3,\hat2,\check2\},\{2,3,\hat1,\check1\},
\{\hat1,\hat2,\hat3\},\{\hat1,\check2,\check3\},
\{\check1,\hat2,\check3\},\{\check1,\check2,\hat3\}\}  , \\
\{F_3\} =\{\{1,2,3,\hat1,\hat2,\hat3,\check1,\check2,\check3\}\}  .
\end{gather*}
Here, we see that $i\perp\hat i$ and $i\perp\check i$, but the realization
in $\R^3$ actually enforces $i\perp j$ as well.

The task then is to classify all connected simple orthogonal $\R$-vector
matroids for given data~$(n,p)$. There are tables in the literature which,
however, do not select for orthogonality. Another disadvantage is the fact
that matroids capture linear dependencies of covector subsets at any rank
up to~$r$, while the WDVV equation sees only coplanarities. Therefore,
it is enough to write down only the 2-f\/lats, which brings us back to the
complete irreducible linear simple hypergraphs again.
Still, the advantage of matroids over hypergraphs is that they provide a~natural setting for the orthogonality property and the partial-isometry
condition~(\ref{pariso}). Once we have constructed a parametric
representation of an $\R$-vector matroid as a family of
$n \times p$~matrices~$A$, we may implement the orthogonalities and directly
test~(\ref{pariso}) for all nontrivial planes~$\pi$. A~{\it good matroid\/}
is one which passes the test and thus yields a (family of) solution(s)
to the WDVV equation.

Another bonus is the possibility to reduce a good matroid to a smaller good
one by graphical methods. The two fundamental operations on a matroid~$M$ are
the deletion and the contraction of an element~$a\in\{1,\ldots,p\}$
(corresponding to a covector). In the geometrical representation these
look as follows
\begin{itemize}\itemsep=0pt
\item {\it deletion\/} of $a$,   denoted $M\backslash\{a\}$:
remove the node~$a$ and all minimal $d$-f\/lats it is part of
\item {\it contraction\/} of $a$, denoted $M/\{a\}$:
remove the node $a$ and identify all nodes on a line with~$a$,
then remove the loops and identify the multiple lines created
\end{itemize}
Both operations reduce $p$ by one. Deletion keeps the rank while contraction
lowers it by one. On the matrix~$A$, the former means removing the
column~$a$ (corresponding to the covector~$\a_a$) while the latter
in addition projects orthogonal to~$\a_a$. Connectedness has to be rechecked
after deletion, but simplicity and the $\R$-vector property are hereditary
for both actions! Furthermore, contraction preserves the orthogonality,
but deletion may produce a non-orthogonal matroid. Since the contraction
of a good matroid corresponds precisely to the restriction of $\vee$-systems
introduced by~\cite{feives1,feives2}, we are conf\/ident that it generates
another good matroid. A similar statement holds for the multiple deletion
which produces a $\vee$-subsystem in the language of~\mbox{\cite{feives1,feives2}}.

The f\/irst nontrivial dimension is $n = r = 3$, where simple
matroids (determined by $\{F_2\}$) are identical with complete
linear simple hypergraphs (given by $\{H_\mu\}$).
Their number grows rapidly with the cardinality~$p$: \\[2pt]
\begin{tabular}{|l|ccccccccccc|}
\hline
number $p$ of covectors
& 2 & 3 & 4 & 5 & 6 & 7  &  8 &   9 &   10 &     11 &       12  \\
\hline
how many simple matroids? $\phantom{\Big|}$
& 0 & 1 & 2 & 4 & 9 & 23 & 68 & 383 & 5249 & 232928 & 28872972  \\
of the above are connected $\phantom{\Big|}$
& 0 & 0 & 0 & 1 & 3 & 12 & 41 & 307 & 4844 & 227612 & 28639649  \\
of the above are $\R$-vector $\phantom{\Big|}$
& 0 & 0 & 0 & 1 & 3 & 11 & 38 & ? & ? & ? & ?  \\
of the above are orthogonal $\phantom{\Big|}$
& 0 & 0 & 0 & 0 & 1 & 1  & 1 & 1 & 3 & ? & ?  \\
\hline
\end{tabular}
\\[4pt]
Below, we list all good ($\checkmark$) and a few bad ($\lightning$) cases
up to $p = 10$, with graphical/geometric representation and the name of the
corresponding root system. Parameters $s,t,u$ indicate continuous moduli.\\[4pt]
$\{\{123\},\{145\}\}$ \quad\ \
\begin{minipage}{1cm}\includegraphics[width=1cm]{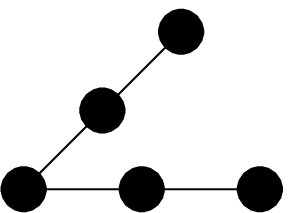}\end{minipage}
\qquad\qquad\qquad\qquad\ \ $A_3\backslash\{6\}$ \hfill
$\R$-vector but not orthogonal\ $\lightning$ \\[2pt]
$\{\{123\},\{1456\}\}$ \qquad\quad\ \
\begin{minipage}{1.5cm}\includegraphics[width=1.5cm]{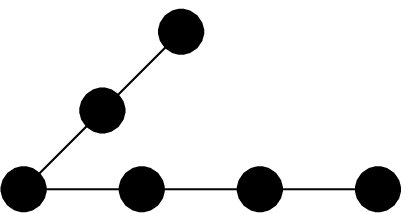}\end{minipage}
\qquad\qquad\qquad $B_3\backslash\{4,5,9\}$ \hfill
$\R$-vector but not orthogonal\ $\lightning$ \\
$\{\{123\},\{145\},\{356\}\}$ \quad
\begin{minipage}{1cm}\includegraphics[width=1cm]{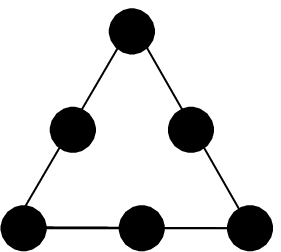}\end{minipage}
\qquad\qquad\qquad $D(2,1;\a)\backslash\{7\}$ \hfill
$\R$-vector but not orthogonal\ $\lightning$ \\
$\{\{123\},\{145\},\{356\},\{246\}\}$ \quad
\begin{minipage}{1cm}\includegraphics[width=1cm]{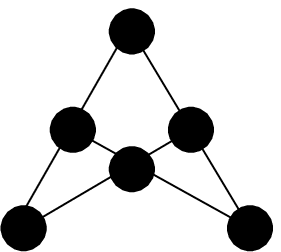}\end{minipage} =
\begin{minipage}{1cm}\includegraphics[width=1cm]{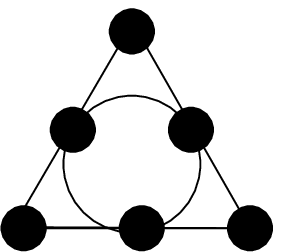}\end{minipage}
\hfill $A_3(s,t,u)$\ $\checkmark$ \\
$\{\{123\},\{145\},\{356\},\{347\},\{257\},\{167\}\}$ \quad
\begin{minipage}{1cm}\includegraphics[width=1cm]{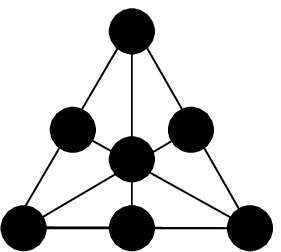}\end{minipage}
\hfill $\underline{6}\oplus\underline{4}\ \textrm{of}\ A_3\ =\ \
D(2,1;\alpha)(s,t)$\ $\checkmark$ \\
$\{\{123\},\{145\},\{356\},\{347\},\{257\},\{167\},\{246\}\}$
\quad
\begin{minipage}{1.3cm}\includegraphics[width=1.2cm]{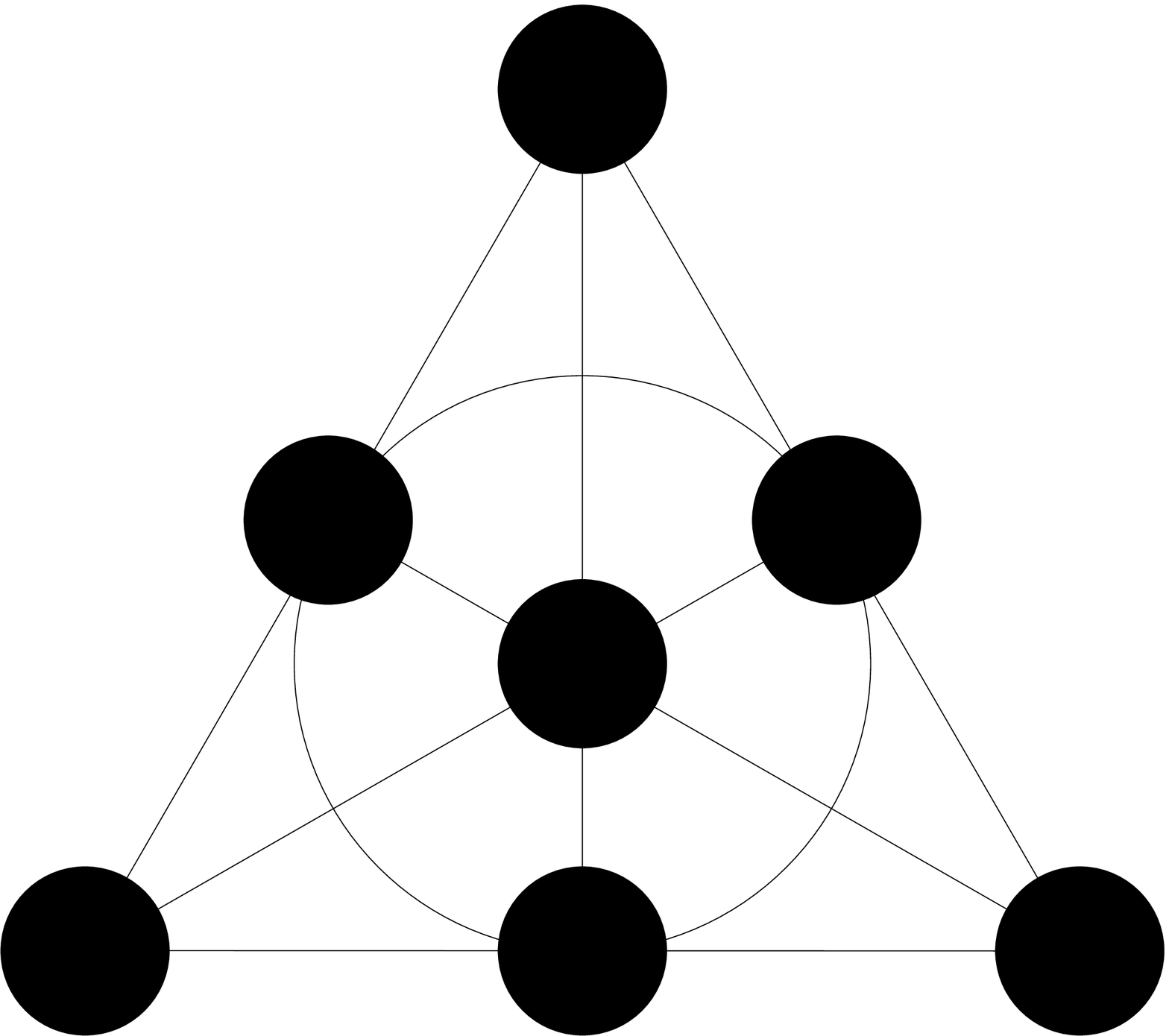}\end{minipage}
\hfill Fano matroid -- not $\R$-vector\ $\lightning$ \\
$\{\{123\},\{145\},\{356\},\{347\},\{257\},\{248\},
\{1678\}\}$ \quad\qquad
\begin{minipage}{1.4cm}\includegraphics[width=1.3cm]{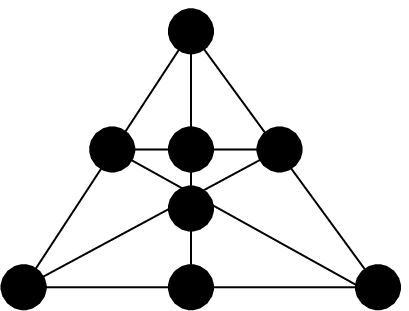}\end{minipage}
\hfill $B_3\backslash\{9\}(s,t)$\ $\checkmark$ \\
$\{\{123\},\{145\},\{347\},\{257\},
\{2489\},\{1678\},\{3569\}\}$ \quad\qquad\qquad
\begin{minipage}{1cm}\includegraphics[width=1cm]{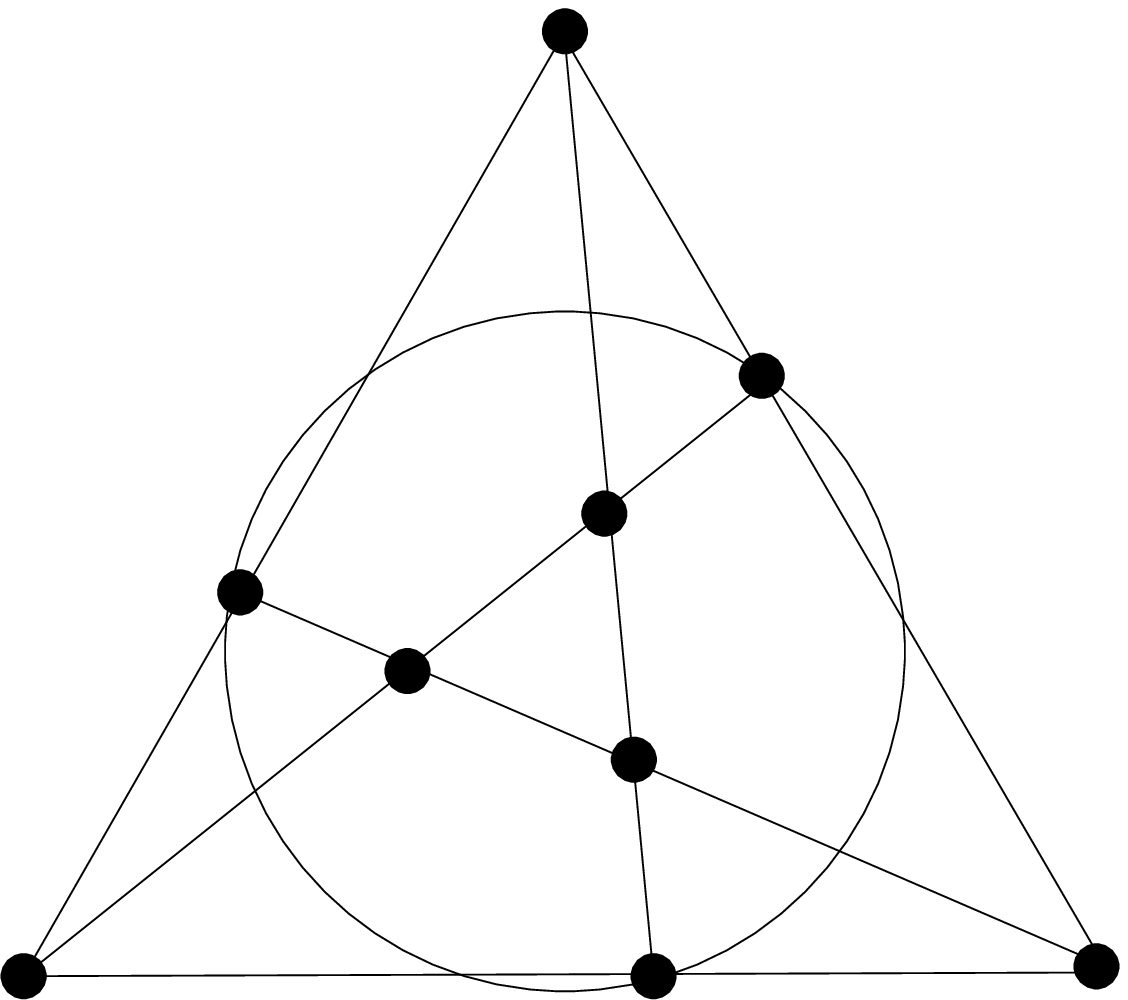}\end{minipage}
\hfill $B_3(s,t,u)$\ $\checkmark$ \\[4pt]
$\{\{150\}\{167\}\{259\}\{268\}\{456\}\{479\}
\{480\}\{1234\}\{3578\}\{3690\}\}$ \
\begin{minipage}{1cm}\includegraphics[width=1cm]{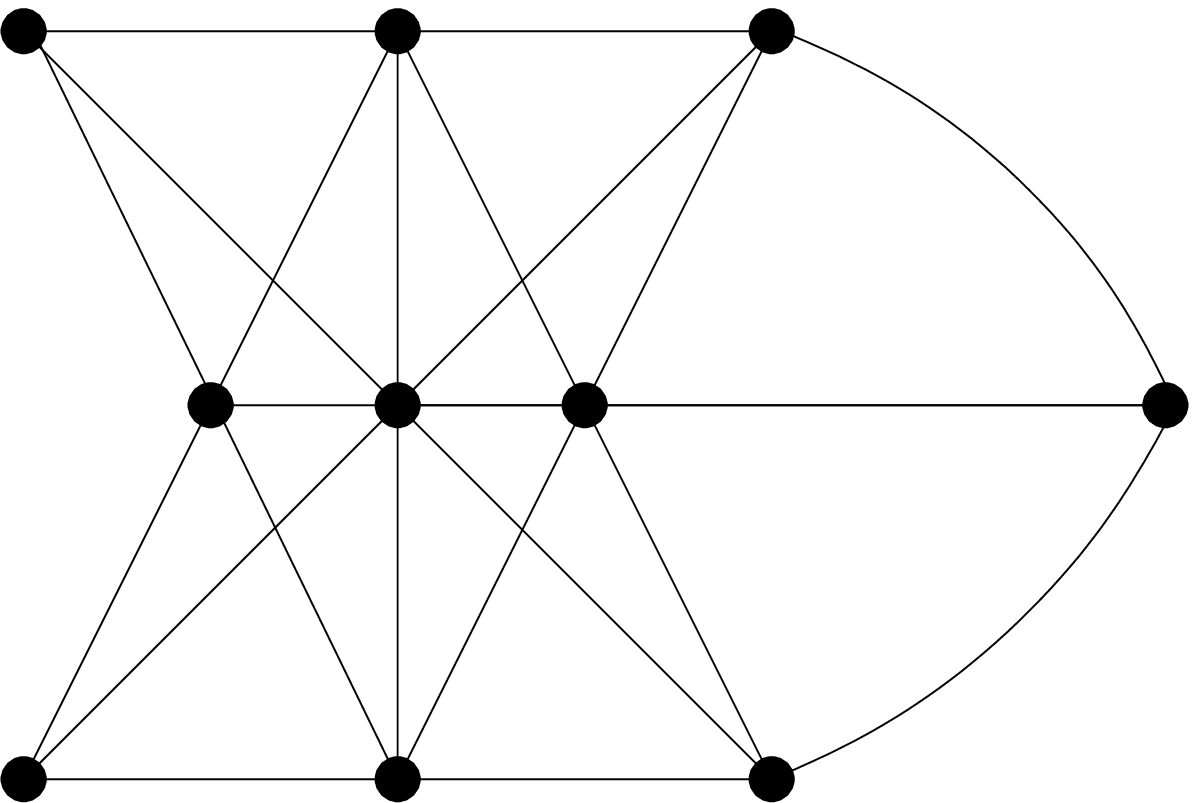}\end{minipage}
\hfill $\subset AB(1,3)(t)$\ $\checkmark$ \\[4pt]
$\{\{179\}\{289\}\{356\}\{378\}\{457\}\{468\}
\{490\}\{1234\}\{1580\}\{2670\}\}$ \
\begin{minipage}{1cm}\includegraphics[width=1cm]{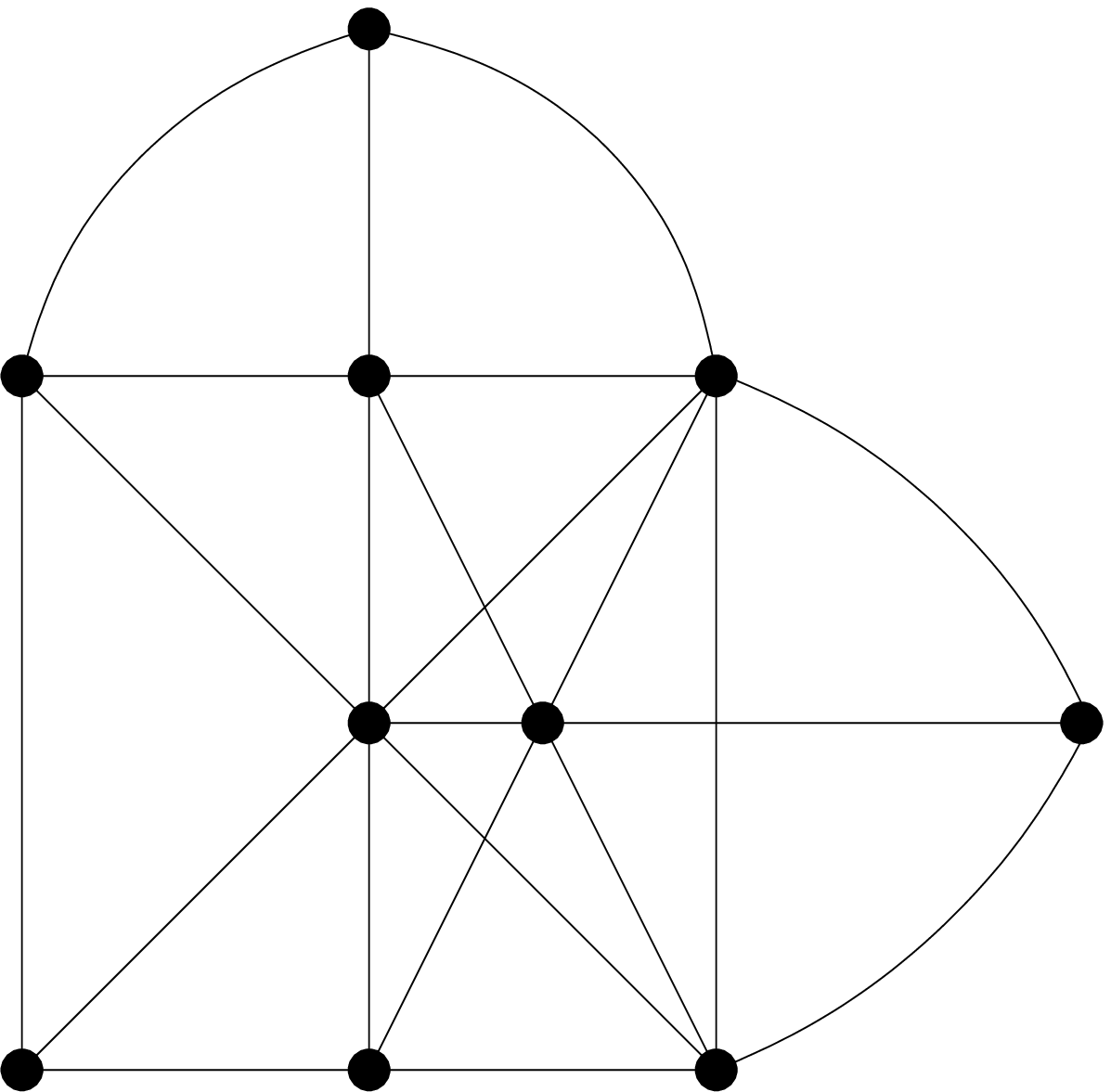}\end{minipage}
\hfill $\subset AB(1,3)(t)$\ $\checkmark$

\noindent
The last two lines (with $p = 10$) arise from restrictions of a one-parameter
deformation of the $p = 18$ exceptional Lie superalgebra $AB(1,3)$ root
system~\cite{feives1}. More precisely, the f\/irst of these two cases is rigid
and can also be obtained from the $E_6$ roots, while the second case retains
the deformation parameter.

We have developed a Mathematica program which automatically generates
all hypergraphs subject to the simplicity, linearity, completeness and
irreducibility properties up to a given~$p$. Furthermore, hypergraphs that
admit no orthogonal covector realization are ruled out, thereby
drastically reducing their number.
For a generated hypergraph we then gradually build a~pa\-ra\-met\-ri\-za\-tion of
the most general admissible set of covectors whereby it turns out
whether the hypergraph is representable. A major step forward would be
to completely automate this process also; we are conf\/ident that this is feasible.
Finally, on the surviving families~$A(s,t,\ldots)$ of covector sets,
the program tests the partial-isometry property~(\ref{pariso})
equivalent to the WDVV equation, for all nontrivial planes~$\pi$.

\begin{wrapfigure}[7]{r}{4.5cm}
\vspace{-2mm}
\includegraphics[width=4.5cm]{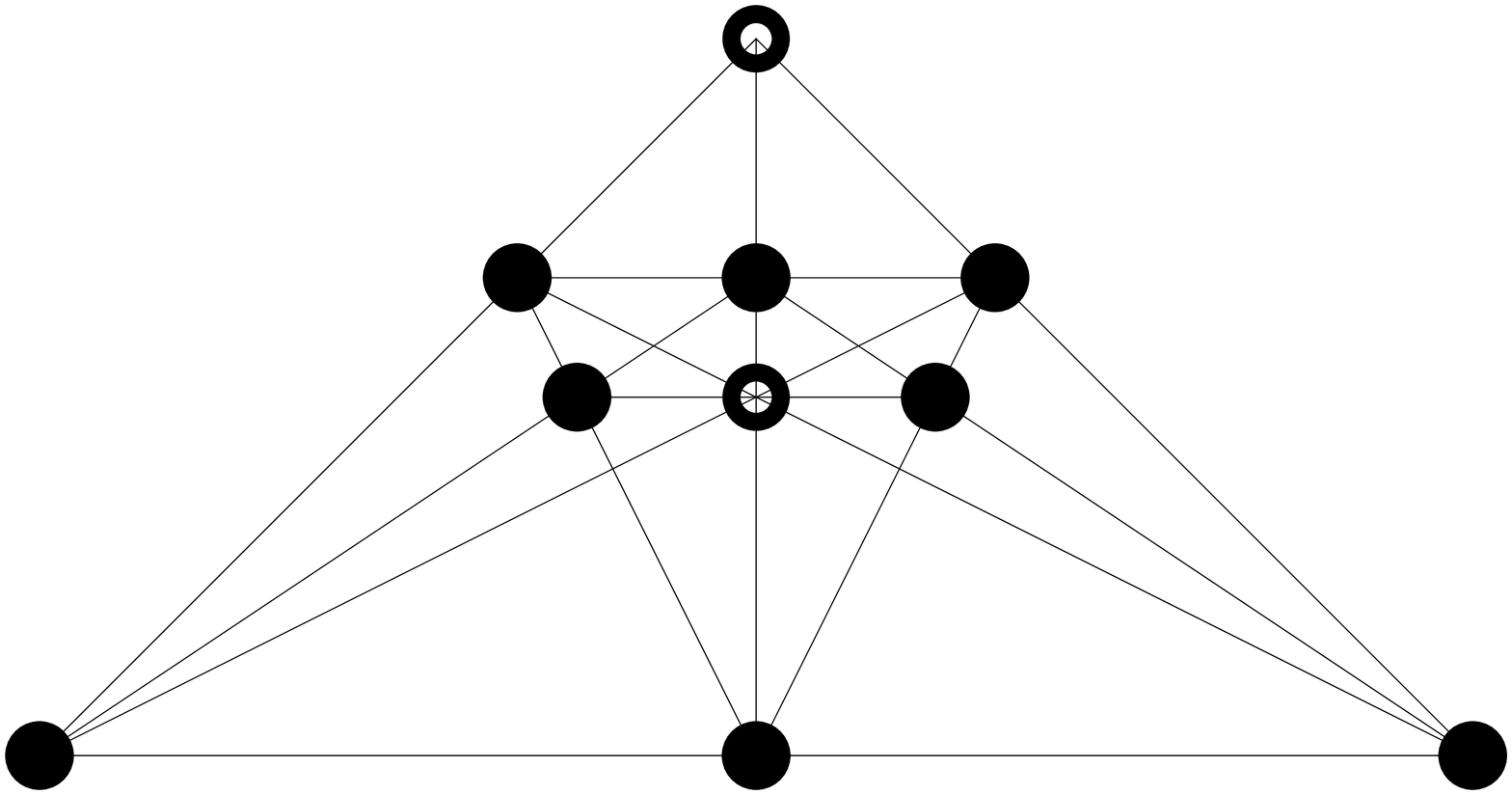}
\end{wrapfigure}
A natural conjecture is that our class of hypergraphs or matroids always
produces WDVV solutions, rendering this f\/inal test obsolete.
However, running the program for a while reveals a counterexample at
$(n,p)=(3,10)$, given by the hypergraph to the right.
In this diagram, the hollow nodes indicate additional orthogonality
inside a plane spanned by four covectors.
We must conclude that a geometric construction of WDVV solutions is
still missing.

Although the connected simple orthogonal $\R$-vector matroids are not
classif\/ied and the WDVV property does not automatically follow from
such a matroid, this approach is still useful in exhausting all covector
solutions for a small number of covectors at low dimension, i.e.\ for
a~limited number of particles. In this way one of us has, in fact,
proven~\cite{schwerdtfeger,notebook} that there are no other four-particle solutions
($n = 3$) with $p\le10$ beyond those determined in~\cite{feives1,feives2}.
The matroid itself does not capture the moduli space of solutions
with a given linear dependence structure, but its systematic realization
by an iterative algorithm will do so (as it did for $n = 3$).
Around a given solution, the local moduli space may be probed by investigating
the zero modes of the WDVV equation linearized around it.

\section{Summary}

We begin by listing the main points of this article:
\begin{itemize}\itemsep=0pt

\item
$\cN{=} 4$ superconformal $n$-particle mechanics in $d = 1$ is
governed by $U$ and $F$
\item
$U$ and $F$ are subject to inhomogeneity, Killing-type and WDVV conditions
\item
a geometric interpretation via f\/lat superpotentials
gave new variants of the integrability 
\item
there is a structural similarity to f\/lat and exact Yang--Mills connections
\item
the general 3-particle system is constructed, with 
three couplings and one free function
\item
higher-particle systems exist, tedious to construct;
hypergeometric functions appear
\item
the covector ansatz for $F$ leads to partial isometry conditions
with multipliers $\lap$
\item
f\/inite Coxeter root systems and certain deformations thereof
yield WDVV solutions
\item
certain solution families admit an ortho-polytope interpretation
\item
hypergraphs and matroids are suitable concepts for a classif\/ication
of WDVV solutions
\item
the generation of candidates can be computer programmed
\item
not all connected simple orthogonal $\R$-vector matroids are `good'
\end{itemize}

There remain a lot of open questions.
First, can our hypergraph/matroid construction program detect new
WDVV solutions not already in the list of~\cite{feives1,feives2}?
Second, given a `good' matroid, can we generate its moduli space,
e.g.\ by linearizing the WDVV equation around it?
Third, the explicit Hamiltonian of the $\cN{=} 4$ four-particle Calogero
system is still unknown.
Fourth, can one construct $\bfu$ as a path-ordered exponential of $\diff\bff$
in a practical way?
Fifth, what happens if we allow for twisted superf\/ields in the
superspace approach?
We hope to come back to some of these issues in the future.

\subsection*{Acknowledgements}

The authors are grateful to Martin Rubey for pointing them to
and helping them with hypergraphs and matroids. Of course, all
mistakes are ours!
O.L.\ acknowledges fruitful discussions with Misha Feigin, Evgeny Ivanov,
Sergey Krivonos, Andrei Smilga and Sasha Veselov.
He also thanks the organizers of the Benasque workshop for a wonderful job.

\pdfbookmark[1]{References}{ref}
\LastPageEnding

\end{document}